\nonstopmode
\documentclass{article}
 
\usepackage{arxiv}

\usepackage{apacite}
\usepackage[utf8]{inputenc} % allow utf-8 input
\usepackage[T1]{fontenc}    % use 8-bit T1 fonts
\usepackage{hyperref}       % hyperlinks
\usepackage{url}            % simple URL typesetting
\usepackage{booktabs}       % professional-quality tables
\usepackage{amsfonts}       % blackboard math symbols
\usepackage{nicefrac}       % compact symbols for 1/2, etc.
\usepackage{microtype}      % microtypography
\usepackage{graphicx}
\usepackage{natbib}
\usepackage{doi}
 \usepackage{amsmath}
\usepackage{threeparttable}
\usepackage{caption}
\usepackage{subcaption}
\usepackage{tabularx}
\DeclareUnicodeCharacter{200B}{}  % ignore zero-width space

\title{Assessing Reporting Delays in ACLED Conflict Event Data}

\author{
\begin{tabular}{cc}
\textbf{Faniry A. Razakason\thanks{Corresponding author: faniry.razakason@stat.uni-muenchen.de}} & \textbf{Daniel Racek} \\
Institute of Statistics, LMU Munich & Institute of Statistics, LMU Munich \\
München, Germany & Center for Crisis Early Warning, University of the Bundeswehr Munich \\
 & München, Germany\\
\\
\textbf{Paul W. Thurner} & \textbf{Göran Kauermann} \\
Institute of Political Science, LMU Munich & Institute of Statistics, LMU Munich \\
München, Germany & Munich Center for Machine Learning \\
 & München, Germany \\
\end{tabular}
}

\begin{document}

\maketitle
\begin{abstract}

Timely and accurate conflict event data are essential for real-time monitoring, forecasting, and policy response. Yet near-real-time conflict datasets such as the Armed Conflict Location \& Event Data Project (ACLED) are subject to reporting delays, that is, delays between event occurrence and first inclusion in the database. Such delays can introduce bias in short-term analyses and forecasts. This study provides a statistical analysis of reporting delays for African events recorded in ACLED's weekly releases from June 30, 2024, to June 1, 2025. Treating delay as a discrete time duration, we estimate grouped proportional hazards models with additive-linear and smooth terms incorporating event-level, spatial, and country-level covariates. Our results show that more than half of events are reported within two weeks, but delays vary systematically by event type, fatalities, geographic location, and political regime. Higher-fatality events are reported more quickly, while events in more restrictive political and informational environments tend to be reported more slowly. We also find substantial between-country heterogeneity, and country-specific analyses indicate that event-level effects differ across contexts. These findings show that reporting delays are structured rather than random and that real-time conflict analysis must account for them. More broadly, they provide an empirical foundation for developing nowcasting approaches to correct short-term underreporting in conflict event data.

\end{abstract}


\newpage

\section{Introduction}

Accurate and timely information on conflict events and fatalities is central to monitoring armed violence \citep{dowd2020comparing}, guiding humanitarian responses \citep{paulus2023reinforcing}, and supporting early-warning systems \citep{rod2024review}. However, their effectiveness critically depends on the quality of the datasets used to build them. In recent years, renewed debate has emerged about the quality of conflict event data produced by major providers, most notably ACLED \citep{miller2022agenda} and UCDP \citep{oberg2025measurement}. These debates focus on biases and errors arising from multiple sources, including omission (missing events), inflation (duplicates or false positives), and misrepresentation (erroneous event details). 

%The frequency with which conflict event datasets are updated has increased substantially in recent years. The Uppsala Conflict Data Program (UCDP) \citep{sundberg2013introducing} introduced its Candidate Events Dataset \citep{hegre2020introducing} to provide preliminary monthly updates, while the Armed Conflict Location \& Event Data Project (ACLED) \citep{raleigh2010introducing} now issues weekly releases \citep{ACLED.02.11.2023}. While these innovations allow researchers and practitioners to follow conflicts in near-real time, they also introduce challenges related to delayed and incomplete reporting. Figure \ref{fig:1} illustrates the problem schematically. At timepoint $t$, the lowest (black) line shows the number of events reported in the database. At $t+1$, additional events that had occurred are added (blue line). By $t+2$, the count rises again (green line). Looking solely at $t$, would yield an underestimate of events for that period, as the number of events is updated twice in the timeperiods that follow, and thus would induce a bias.

\begin{figure}[b]
    \centering
      \includegraphics[scale=0.7, trim=0 6in 0 1.4in, clip]{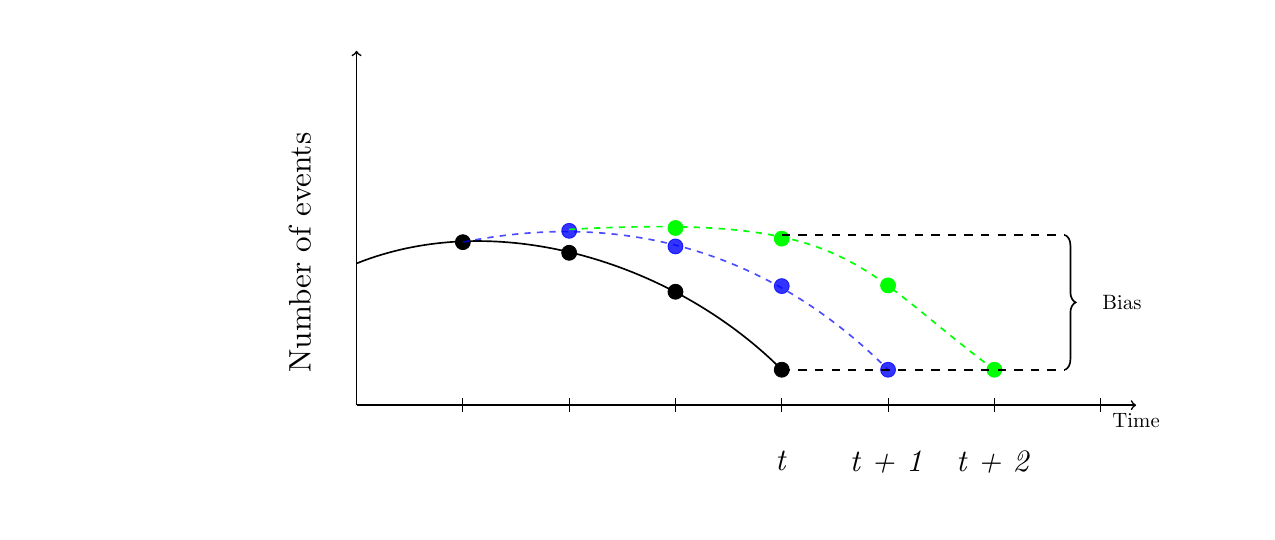}
    \caption{Sketch of reporting delay. The black (bottom) line shows the reported events at time point $t$. The blue (middle) line gives the numbers at $t+1$ and the green (top) line at $t+2$.  }
    \label{fig:1}
\end{figure}

The frequency with which conflict event datasets are updated has increased substantially in recent years. The Uppsala Conflict Data Program (UCDP) \citep{sundberg2013introducing} introduced its Candidate Events Dataset \citep{hegre2020introducing} to provide preliminary monthly updates, while the Armed Conflict Location \& Event Data Project (ACLED) \citep{raleigh2010introducing} now issues weekly releases \citep{ACLED.02.11.2023}. Although these innovations allow researchers and practitioners to follow conflicts in near-real time, they also introduce additional data challenges, which are related to potentially incomplete reporting. In these higher-frequency datasets, there is typically a delay between when an event actually occurs and when it first appears in the database. In the following, we refer to this as a {\sl reporting delay}. Figure \ref{fig:1} illustrates the problem schematically. At time $t$, the lowest line (black) shows the number of events reported in the database. In $t+1$, additional events that had occurred are added to the database (blue line). By $t+2$, this count rises again (green line). Looking solely at $t$ would yield an underestimate of the events for that period, since the number of events is updated twice in the time periods that follow.%, and thus would induce a bias.

Understanding and quantifying these reporting delays is important for at least three reasons. First, delays in database inclusion can bias short-term analyses and distort inference about recent changes in conflict intensity. Second, real-time forecasts that
 treat the most recent reported counts as complete are likely to underestimate uncertainty and risk producing biased predictions. Third, these biases can mislead decision-makers and contribute to inappropriate or mistimed policy and humanitarian responses.

Although conflict scholars have long debated coverage and accuracy in event data with respect to scope and sourcing \citep{weidmann2015accuracy,raleigh2023political,oberg2025measurement}, systematic study of delayed reporting, and how delays vary with event characteristics, geography, and political context remains limited. \cite{a872ea27-0222-38ba-a5a1-57c7686006ab, donnay2019integrating} provide a preliminary attempt to describe how location, event type, and source influence delays, but treats delay as a binary outcome - instead of a continuous measure - and does neither estimate their relative impact nor quantify uncertainty.

%While conflict research has long debated coverage and accuracy in event data based on dataset scope and sourcing \citep{weidmann2015accuracy,raleigh2023political,oberg2025measurement}, delayed reporting and how these delays vary with event features, space, and political context have barely been systematically studied. \cite{a872ea27-0222-38ba-a5a1-57c7686006ab} offers a descriptive overview of how reporting delays are affected by the location, type, and report source of the event. There is, however, no treatment of uncertainty, and reporting delays were aggregated into a binary event, rather than modeled as quantities that can take different values. 
By contrast, this problem has been well theorized in other domains. For example, in epidemiology, daily case counts of infectious diseases are affected by both incubation and reporting times, leading to delays between infection and registration. The statistical technique to correct for this is known as {\sl nowcasting} and was widely applied during the COVID-19 pandemic \citep{schneble2021nowcasting,de2022regional}. Nowcasting dates back to \cite{zeger1989statistical} and \cite{lawless1994adjustments}, and has been adapted in various fields, including economics (e.g.\ \citealp{giannone2008nowcasting}) and meteorology (e.g.\ \citealp{browning1989nowcasting}, \citealp{shi2015convolutional}). All share a common logic: multiple snapshots of the same time series, each subject to revision, can be used to infer the current but incompletely observed state of the system. The major requirement for applying nowcasting in such scenarios is that the causes for delay and a possible stochastic structure of the delay are well-understood. Conflict event data, which are continuously revised as new reports become available, exhibit precisely this revision structure. Adapting nowcasting techniques to conflict event data requires, first of all, a detailed understanding of the mechanisms that generate reporting delays. This is exactly the objective of this paper.

We present a detailed empirical analysis of (short-term) reporting delays in ACLED's weekly releases for Africa. Using a series of weekly ACLED snapshots collected between June 30, 2024 and June 1, 2025, we observe the week in which each event first appears in the database and define the reporting delay as the number of weeks between occurrence and first appearance. We treat delay as a discrete time-to-event outcome and model it with a grouped proportional-hazards specification embedded in a generalized additive model (GAM) to provide a flexible, smooth baseline hazard and to permit potentially non-linear covariate effects \citep{wood2017generalized}. This approach allows us to quantify how delays depend on event-level attributes, geographic factors, and country-level characteristics. Our results show that delays are not random noise but are systematically influenced by factors which are also discussed in other contributions to the literature on event biases. More importantly, delays also  exhibit substantial between-country heterogeneity. Delays vary across event types and depend on country-level conditions, including political regime, socio-economic context, information accessibility and media censorship, and overall conflict severity. Event-level characteristics - such as local conflict severity and spatial context - also affect reporting timeliness, and these event-level effects differ across countries.

We emphasize that the biases we study impair exclusively real-time monitoring and forecasting. In retrospective analyses using curated data, i.e. once the reporting window has effectively closed, this source of bias is negligible. We also want to note that the identification and modeling of reporting delays require repeated snapshots of the dataset over time so that the first appearance of each event can be observed. Consequently, regular downloads of ACLED releases are necessary (more details are provided below).

The paper proceeds as follows. As we expect that the well-known factors inducing reporting biases more generally also impact reporting delays, we first review existing work on bias and measurement in conflict event data. We then describe the data collection procedure and variable construction, before presenting the statistical framework for modeling reporting delays. Next, we report our main findings. The final section discusses the implications of our results and concludes. 
%For completeness, we emphasize that the specific “bias” we are investigating in this paper is solely important for the objective of real-time monitoring and forecasting. For other analyses, using past data sources, reporting delay is not an issue, as long as the data is "final", meaning all conflicts are included for the time window of the intended data analysis. We also remark that the identification of reporting delays requires the possibility to continuously monitor the insertion of new data, meaning that regular downloads of the data are necessary. More details in this matter are provided in the paper. 
%The paper is organized as follows. In the following, we briefly review the literature on biases in conflict data and summarize its main findings. The subsequent sections describe the data collection process and variable specifications, followed by a detailed presentation of our proposed model for analyzing reporting delays. We then present our main findings. Finally, we discuss the implications of our results and conclude the paper. 
%The remainder of this article is structured as follows. Section 2 explains the data collection process and the construction of weekly snapshots. In Section 3 we thoroughly describe our proposed model to analyse the reporting delays. Then, in Section 4, we present our main findings. Finally, Section 5 discusses the implications of our results and concludes our work. %Section 5 discusses implications for real-time conflict monitoring and outlines how our findings can inform the development of nowcasting tools for conflict data.
\section{Bias in conflict data}

Most quantitative studies in conflict research now rely on disaggregated conflict event data with increasingly fine spatial and temporal resolution. Spatially, conflict events are typically measured at fine-grained geographical levels below the nation-state, either at the PRIO grid cell level of $0.5^\circ \times 0.5^\circ$ (roughly $55 \times 55$ km at the equator) \citep{tollefsen2012prio} or with precise geo-locations \citep{raleigh2010introducing,sundberg2013introducing}. Temporally, the field has likewise shifted from annual aggregates to datasets that record events at a weekly or even daily level. The most timely conflict data are now updated on a weekly basis, either through predominantly human-coded datasets such as the Armed Conflict Location \& Event Data Project (ACLED), the Uppsala Conflict Data Program Georeferenced Event Dataset (UCDP GED), the Global Terrorism Database (GTD), and the Social Conflict Analysis Database (SCAD), or through primarily machine-coded systems such as GDELT and ICEWS/POLECAT. Automated extraction and coding of conflict events has expanded rapidly in recent years \citealp{olsson2020text,hu2022conflibert,croicu2024deep,meher2025conflllama}). Yet efforts to evaluate the validity and reliability of machine-coded data still rely largely on human-coded datasets as the ground truth or benchmark.

A central challenge for conflict research is that measures of political violence are produced through a complex information “supply chain”" that involves reporting, selection, extraction and coding. Each stage introduces potential sources of bias, including selection effects, filtering, and overrepresentation of particular types of events. There has been extensive discussion on whether conflict data are biased in terms of event coverage and geolocation precision (e.g., \citealp{lee2019lost}), as well as with respect to specific event attributes such as death counts. Bias may also arise when actors (e.g. governments, rebel groups, NGOs etc.) have incentives to conceal violence or to strategically exaggerate or downplay certain characteristics of events, including civilian casualties, or to follow unconscious built-in biases in attention assignment. Recently, \cite{gibilisco2023strategic} formalizes these dynamics by developing a strategic theory of under- and over-reporting by governments and non-state actors. Factors that have been identified in the meanwhile extensive literature to cause biases in event data should also induce reporting delays.

The central insight emerging from these contributions is that conflict data are inherently and systematically biased. These biases are not just random noise that can be avoided with larger samples, but are systematically related to structural characteristics such as regime type, information accessibility, the diffusion of communication technologies, and geographic remoteness. For example, if violence is more likely to be reported when it occurs near urban centers or in areas with reliable mobile coverage, statistical models may spuriously attribute conflict onset or intensity to confounding processes such as modernization. In addition, different data sources can produce contradictory conflict histories. Bias can also arise from heterogeneity among observers: journalists, human rights activists, and victims may perceive, record, and prioritize violence differently, as illustrated by recent controversies surrounding reporting on the war in Gaza \citep{alshebli2026mediacoveragewarvictims}. 
%Factors that have been identified in the meanwhile extensive literature to cause biases in event data should also induce reporting delays. This is why we review this literature. We argue that the very same factors also impair the specific process of generating delays and timeliness of reporting.

%The central insight from these contributions is that conflict data are inherently and systematically biased. That is, biases in these data are not merely random noise that can be overcome by large sample sizes, but are systematically related to structural explanatory variables such as regime type, the availability and diffusion of information and communication technologies, or geographic remoteness. For example, if violence is only reported when it occurs near urban centers or in areas with cell phone coverage, statistical models may spuriously attribute conflict onset to confounders such as modernization. Moreover, different sources may yield contradictory conflict histories. Bias can arise from differences among observers: journalists, human rights activists, and victims may perceive, record, and prioritize violence differently, as has recently been evident in the fierce discussions surrounding reporting on the war in Gaza.
This heterogeneity of perspectives is demonstrated by \cite{davenport2002views}. For Guatemala (1977–1995), they compared event lists compiled from newspapers, human rights non-governmental organizations (NGOs), and retrospective eyewitness interviews, revealing different “views of a kill” associated with each type of source. Newspapers reflect an urban and commercial perspective, as media outlets tend to prioritize events that occur in accessible urban settings. As a result, rural violence is systematically underreported due to high access costs, limited audience interest, and also sensitivity to state censorship. By contrast, human rights NGOs adopt an advocacy perspective trying to support marginalized and discriminated populations and mobilizing international attention. This orientation leads them to emphasize events with large casualty numbers, generating a severity bias, while their reporting often lags and depends on the availability of local networks. Eyewitness interviews, in turn, provide a predominantly rural and victim-centered perspective. Extending this argument, \cite{dawkins2021problem} shows that reporting during the South Sudanese civil war was further shaped by predefined external narratives that influenced which events were documented.

%This heterogeneity of perspectives was demonstrated by \cite{davenport2002views}. For the case of Guatemala (1977–1995), they compared event lists based on newspapers, human rights non-governmental organizations (NGOs), and retrospective eyewitness interviews. Their analysis reveals distinct “views of a kill” depending on the source type. Newspapers reflect an urban and commercial lens, as media outlets and their personnel tend to favor events occurring in urban environments. They systematically underreported rural violence due to high accessibility costs and limited audience interest in remote indigenous areas, and they were also particularly sensitive to state censorship. By contrast, human rights NGOs adopt an advocacy lens, focusing on their role in supporting disadvantaged populations and mobilizing international attention. As a result, they tend to highlight events with large numbers of casualties, producing a severity bias. At the same time, their data collection often lagged and is dependent on the presence of local networks. Eyewitness interviews, in turn, reflect a victim-centered and rural lens, providing yet another distinct perspective. Investigating the South Sudanese civil war, \cite{dawkins2021problem} further argues that events may be selected according to predefined external narratives.
While \cite{davenport2002views} focus primarily on \textit{who} reports violence, subsequent research has rigorously examined \textit{where} and \textit{how} reporting biases arise. The spatial dimension is particularly consequential for disaggregated conflict studies, which link events to local covariates such as terrain, weather, or poverty. \cite{weidmann2015accuracy} examines the accuracy of reported event characteristics—especially location and severity—once an event enters the news. Comparing media-based event data from UCDP-GED with military “Significant Activity” (SIGACT) reports from Afghanistan, Weidmann validates media reporting against a benchmark. His findings provide clear evidence of a remoteness bias: events that occur in rural or highly insecure areas exhibit greater spatial uncertainty. Reports in such contexts often identify only districts or provinces, rather than specific villages. Weidmann quantifies this degradation in precision, showing that, while district-level patterns are captured reasonably well, locational accuracy declines sharply at finer spatial scales. Consequently, analyzes relying on very fine-grained units, such as small grid cells or village-level data, risk overstating spatial precision. At the same time, Weidmann finds that the reported casualty counts are comparatively robust.

A specific manifestation of reporting bias concerns the interaction between communication technology and the reporting of violence. \cite{pierskalla2013technology} argue that the expansion of mobile phone coverage in Africa facilitated collective action and coordination, thus increasing the incidence of violent events. \cite{weidmann2016closer} challenges this interpretation, suggesting that the observed association may instead reflect reporting bias: mobile phones make it easier to transmit information to journalists or enable journalists to contact witnesses more easily, increasing the number of recorded events in connected areas even if the underlying violence does not increase. \cite{hollenbach2017re} respond by reassessing this claim using sensitivity analyses and simulation-based approaches to test the robustness of the technology–violence relationship to potential reporting bias. Drawing on \cite{dafoe2015cell}, they argue that reporting bias is likely non-linear: small, low-fatality events are more likely to be missed in unconnected areas, whereas large, high-fatality events tend to be reported regardless of connectivity. If the cell coverage effect were purely artifactual, it should vanish when restricting the analysis to high-casualty events. However, they find that the effect remains stable across severity thresholds. In addition, they simulate hypothetical under-reporting by injecting “fake” events into uncovered areas. Even when assuming that up to 25\% of events in such areas are missing, their models continue to recover a positive relationship between mobile coverage and violence, suggesting that this signal is robust to substantial reporting noise.

A related measurement challenge arises from false positives in both machine-coded and human-coded conflict data, which generate “spurious events”, that is, records of interactions that never occurred. \cite{fritz2023all} show that standard Relational Event Models (REMs) are particularly sensitive to such errors because they implicitly treat every recorded event as a true realization of the underlying process. To address this problem, they propose a Relational Event Model for Spurious Events (REMSE), which models the observed event stream as a mixture of two latent processes: a substantive signal process driven by covariates such as rivalry or reciprocity, and a noise process capturing measurement error. Using a Bayesian data augmentation approach, the model estimates the probability that any given event is genuine or spurious. Applied to data from the Syrian civil war, Fritz et al. demonstrate that failing to account for spurious events can lead to biased coefficient estimates. 

Despite this extensive literature on various forms of reporting biases, systematic analysis of reporting delays remains limited. \cite{a872ea27-0222-38ba-a5a1-57c7686006ab} provide an initial exploration of the probable determinants of reporting delays, showing that delay varies with event location, event type, actors, and source characteristics. However, these factors are examined separately rather than jointly, and delay is treated as a binary outcome (delayed versus not delayed) rather than as a duration with an underlying distribution. Beyond this contribution, work that directly studies reporting delays is virtually non-existent. This gap likely reflects the fact that regular, timely update cycles for conflict event datasets have only recently become common. We address this gap by statistically modeling reporting delays and studying their distribution.

\section{Data \& Variable specification}

Our analysis is based on data from the Armed Conflict Location \& Event Data Project (ACLED). To our knowledge, ACLED constitutes the only human-coded global conflict event dataset that is updated on a weekly basis, making it particularly valuable for near-real-time analyses. To measure reporting delays, we downloaded updated versions of the data set every Sunday between June 30, 2024, and June 1, 2025. These weekly snapshots allow us to identify two key time points for each event: the date of event occurrence and the date of first inclusion in the dataset.\footnote{Although each event record contains a timestamp attribute, it is updated whenever any variable is modified. It therefore cannot be used to recover the date on which the event first entered the database.}

Because each ACLED release includes the full historical record dating back to 1997, working with and processing the entire dataset would be computationally intensive. We therefore restrict our analysis to events in Africa, where armed conflict is prevalent and many research studies have been conducted in \citep{von2016civil,bagozzi2017droughts,abidoye2021income,maconga2023arid,racek2024integrating}. Within Africa, we exclude countries with very limited conflict activity to ensure that country-level covariates can be reliably estimated. Specifically, we omit eight countries with fewer than ten recorded conflict events in 2023. Additional details on country selection are provided in the Appendix.

We define reporting delay as the number of weeks between the occurrence of an event and its first appearance in ACLED. To capture newly reported events, we compare consecutive weekly downloads and identify records that appear for the first time in each release. This approach allows us to identify the timing of initial reporting, even in the presence of subsequent updates and corrections. A detailed description of the data construction process, including identification rules and the handling of multiple updates, is provided in the Appendix.

Figure~\ref{fig:plot1} illustrates the reporting delays introduced schematically earlier using ACLED data. The figure visualizes the number of events that occurred between January 1 and June 24, 2024, as recorded in datasets downloaded on June 30, in the subsequent four weeks, and again on December 24, 2024. The x-axis denotes the first day of each week (e.g., “01-01-2024” covers events from January 1 to 7). As shown, the reported number of events for any given week increases between download dates. The difference between the initial June 30 dataset and subsequent releases reflects newly reported events, whose delays are measured as the interval between occurrence and first inclusion. For instance, the week of June 17, 2024, initially shows 445 events in the June 30 dataset, but later increases to 870 in the December version, thus highlighting substantial reporting delays in ACLED.

\begin{figure}[h]
    \centering
    \includegraphics[width=0.9\textwidth]{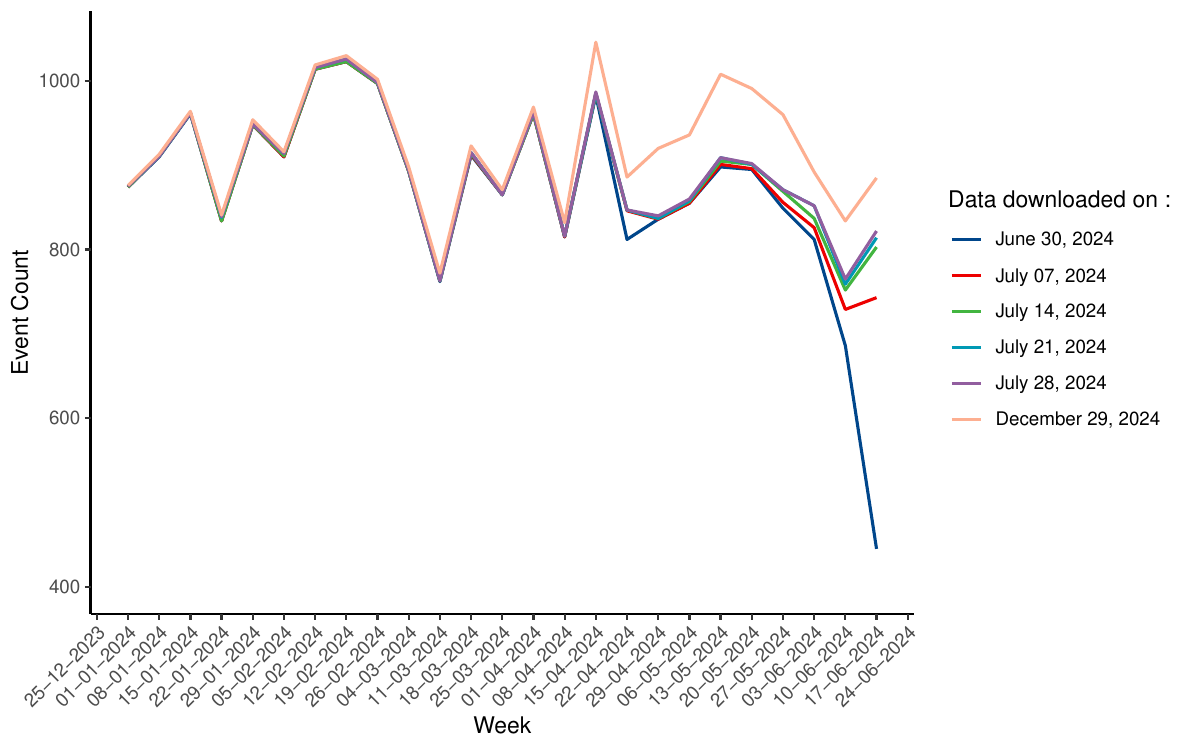} 
    \caption{Reported conflict events over time based on dataset download dates. The figure illustrates the number of events occurring between January 1 and June 24, 2024, as recorded in datasets downloaded on June 30, 2024, the following 4 weeks, and on December 24, 2024}
    \label{fig:plot1}
\end{figure}

\begin{figure}[h]
    \centering
    \begin{subfigure}[t]{0.7\textwidth}
        \centering
        \includegraphics[width=\textwidth]{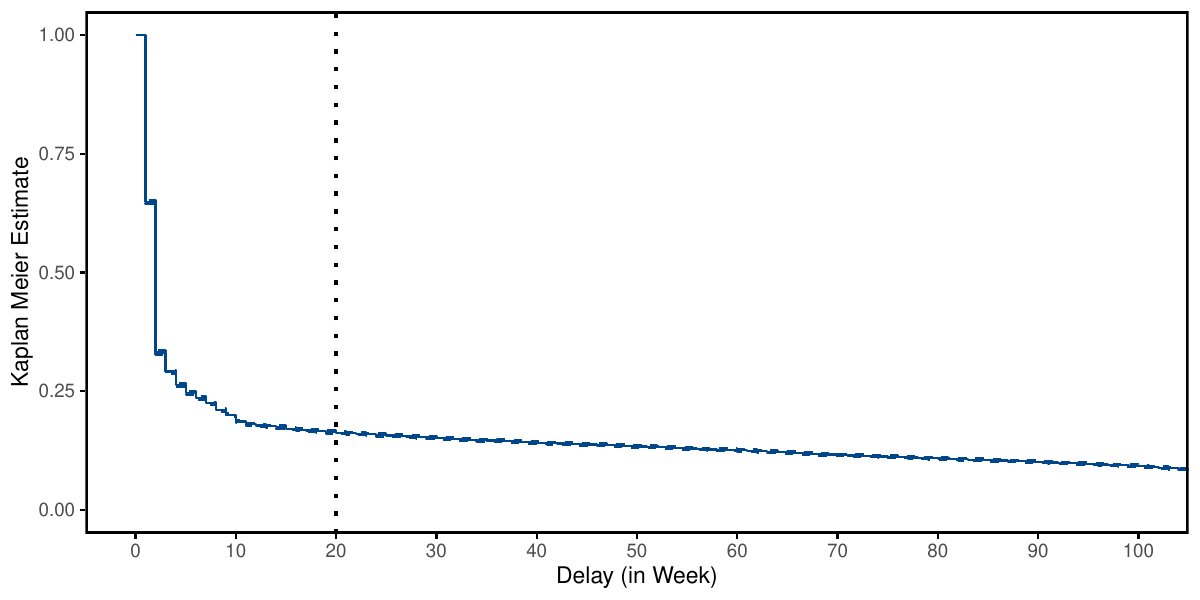}
        \caption{Kaplan--Meier curves of reporting delays (in weeks) for the first 100 weeks.}
        \label{fig:plot3a}
    \end{subfigure}

    \vspace{0.5cm}

    \begin{subfigure}[t]{0.7\textwidth}
        \centering
        \includegraphics[width=\textwidth]{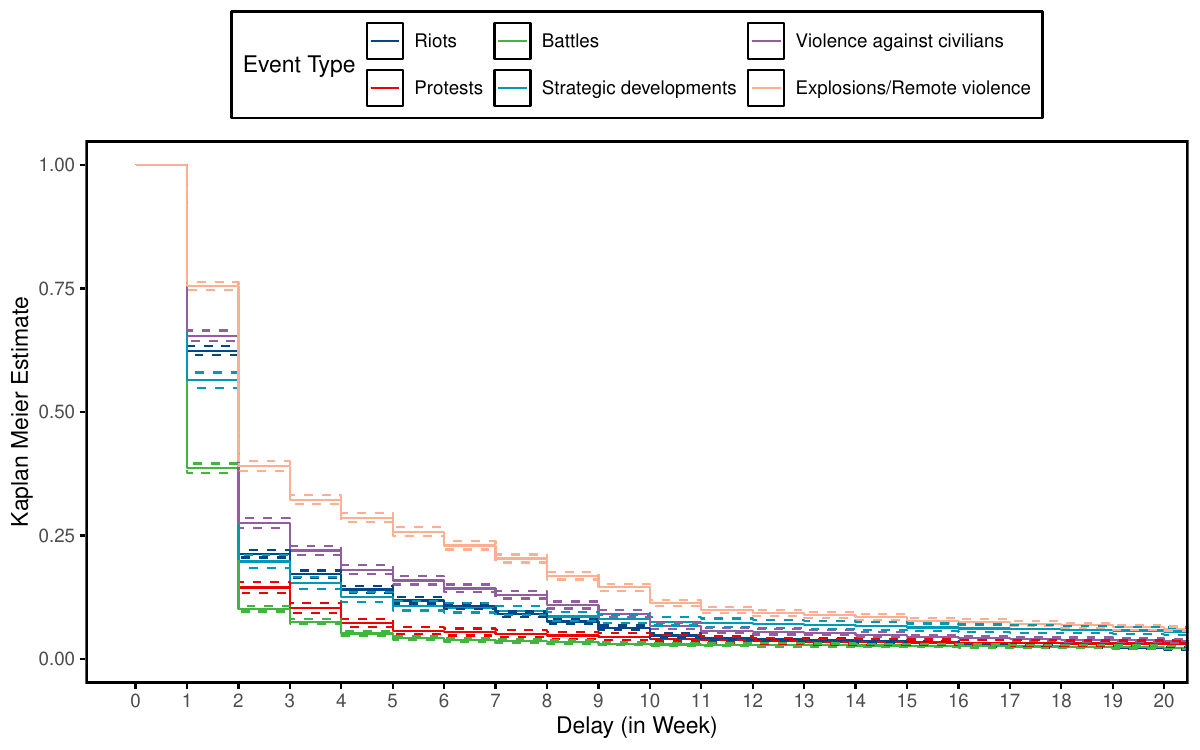}
        \caption{Kaplan--Meier curves of reporting delays (in weeks) for the first 20 weeks, disaggregated by event type.}
        \label{fig:plot3b}
    \end{subfigure}

    \caption{Kaplan--Meier curves of reporting delays: (a) first 100 weeks; (b) first 20 weeks by event type.}
    \label{fig:plot3}
\end{figure}

Treating reporting delay as duration time enables the use of standard survival analysis tools, such as Kaplan–Meier curves. Figure~\ref{fig:plot3a} shows the Kaplan–Meier estimate of reporting delays (in weeks) during the first 100 weeks.
The curve reveals a rapid decline in the probability of reporting during the first 20 weeks, followed by a much slower decrease thereafter. Approximately 84\% of events are reported after 20 weeks, and approximately 90\% after 100 weeks. 
%Approximately 16\% of events are reported after 20 weeks of delay and only 10\% after 100 weeks. 
The steep initial drop indicates that the majority of events enter the dataset within the first few months after occurrence, motivating our focus on this period in the subsequent analysis. The gradual and slow decline beyond 20 weeks is largely driven by historical “bulk uploads”, i.e. instances in which many older events for a given country are added on the same day. We exclude such historical entries (see Appendix for details). After this exclusion, most events are reported within 20 weeks, which we adopt as the censoring threshold in the analyses that follow.

\begin{figure}[ht]
    \centering
    \includegraphics[width=0.7\textwidth, trim=0 1in 0 1.4in, clip]{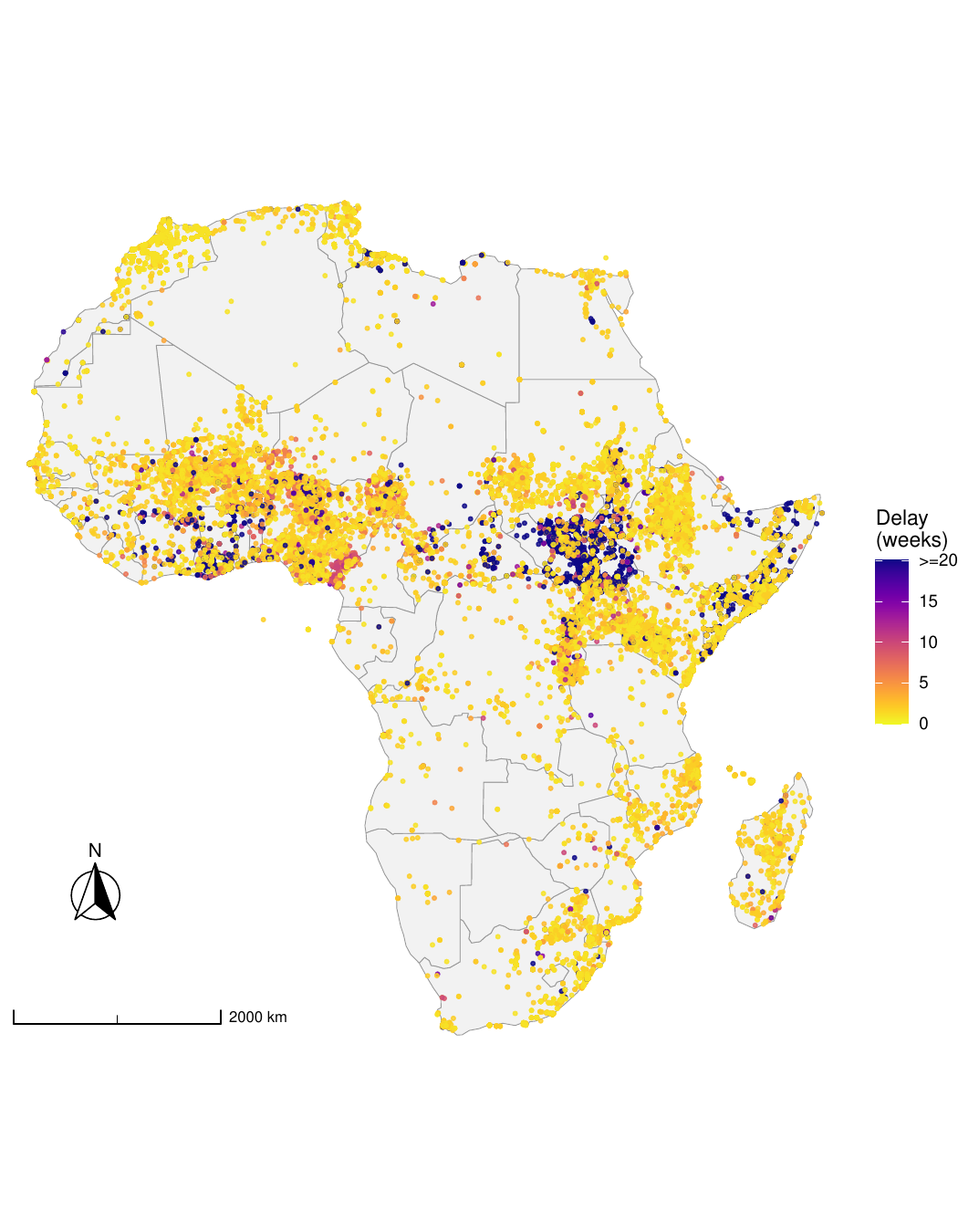} 
    \caption{Conflict events plotted over a map of Africa, with each point colored according to its reporting delay in weeks.}
    \label{fig:plot_africa}
\end{figure}

Figure~\ref{fig:plot3b} disaggregates the Kaplan–Meier curves by event type after excluding historical records. The figure reveals systematic variation across event categories. Explosions and remote violence, for example, tend to exhibit longer reporting delays in the first 10 weeks, whereas battles and protests are typically reported more quickly. Figure~\ref{fig:plot_africa} provides a spatial perspective on these delays. Each point represents a conflict event in Africa and is coloured by its reporting delay (in weeks). The map indicates substantial spatial heterogeneity, with longer delays more common in parts of Central and West Africa, while reporting is generally more timely in other regions, particularly Southern Africa. These patterns suggest that timeliness is shaped by both cross-country and within-country contextual factors.

Motivated by these observations, we investigate the determinants of reporting delays using a parsimonious set of theoretically motivated event-level and contextual covariates. 
%by drawing on both event-specific and contextual variables. Rather than attempting to model every possible influence, we focus on a concise set of theoretically motivated variables designed to uncover systematic patterns in the delay process. 
From ACLED, we use the type of event and the number of fatalities as well as the event geolocation. Previous work suggests that the severity of the event is an important determinant of underreporting. In particular, \cite{croicu2022reporting} shows that non-fatal events are substantially more likely to go unreported. There is also evidence that the type of event shapes the attention of the media. For example, \cite{kearns2019some} shows that certain terrorist incidents receive disproportionately greater coverage.

Beyond event-level characteristics, we incorporate several external indicators that capture the broader political, socioeconomic, and spatial contexts in which reporting occurs. First, we consider regime type, measured using the v2x\_regime classification introduced by \citet{luhrmann2018regimes} from the Varieties of Democracy (V-Dem) Project \citep{coppedge2024v}.\footnote{The classification distinguishes four regime types: closed autocracy, electoral autocracy, electoral democracy, and open democracy. Because the analysis focuses on African countries, only the first three categories are observed in the data.} Regime type provides a summary measure of political freedom and institutional constraints. Democratic systems are typically associated with greater transparency of information, greater freedom of the press, and greater public accountability \citep{hollyer2011democracy}, all of which may facilitate more timely reporting. Moreover, \cite{hoiby2019journalism} emphasizes how pressure on journalism can result in underreporting of conflict events. For this reason, we include censorship-related variables to capture the effects of information control and include internet access as an additional indicator of information availability. Specifically, we use government censorship of the media (v2mecenefm), self-censorship of the media (v2meslfcen) and internet access (v2mecenefibin) from V-Dem \citep{coppedge2024v, pemstein2018v}.

We further account for socioeconomic factors. GDP per capita \citep{WorldBank_GDPperCapita} is included as a proxy for both state capacity and the quality of information infrastructure. Wealthier countries tend to have more developed communication networks and monitoring systems \citep{pradhan2018information}, which may accelerate the reporting of conflict events. In addition, we include total population to control for scale effects, given evidence that population size increases the risk of conflict \citep{raleigh2009population, bruckner2010population}, we include total population to control for its potential influence on reporting capacity.

Finally, to capture local reporting conditions, we calculated a local population exposure measure defined as the total population living within 50 km of each event location, using gridded population data of 1 km\textsuperscript{2} from WorldPop \citep{WorldPop2016}. This variable allows us to assess whether events that occur in more densely populated areas are reported more promptly. A 50 km radius is chosen to account for uncertainty in event locations (see, e.g., \citealp{weidmann2015accuracy}).\footnote{For each event, we construct a circular buffer with a 50 km radius and sum the population values of all 1~km~$\times$~1~km cells whose centroids fall within the buffer.} We also include the geodesic distance to the national capital and to the nearest international border as measures of remoteness, capturing differences in monitoring capacity and media reach between central and more remote regions, as well as strategically sensitive areas such as border zones. All external covariates are taken from the most recent data releases: 2024 for GDP per capita and all the data from V-Dem, and 2020 for population.

\section{Methods}
In the following, we introduce the statistical framework used to analyze the reporting delays. Let $T_i$ denote the date of occurrence of the event $i$, and let $R_i$ denote the date on which it first appears in the database. The reporting delay is then defined as $D_i = R_i - T_i$, with $R_i > T_i$. Because both \( T_i \) and \( R_i \) are calendar dates, \( D_i \) measures the elapsed time. We express \( D_i \) in weeks, reflecting both ACLED's weekly update cycle and our data collection frequency. The minimum delay is therefore one week, corresponding to an event reported in the first possible release. To exclude historical anomalies with very long delays, we censor observations at \( D_{max} = 20 \), informed by the flattening of the delay distribution beyond 20 weeks (see Figure~\ref{fig:plot3}). Each event \( i \) is associated with a set of covariates that form the basis for two complementary model specifications designed to capture heterogeneity in reporting delays in an interpretable manner. Both models share the same statistical framework for modeling delay and differ only in the covariates included.

We proceed in two steps. First, we estimate a \textbf{country-level} model (M1) to assess whether reporting delays differ systematically between countries. Model M1 incorporates a set of national-level covariates intended to capture structural, demographic, and institutional conditions that may influence the timeliness of reporting. Specifically, the model includes the logarithm of GDP per capita, total population, and the cumulative fatalities associated with the relevant sub-event in a country from January 1, 2024, to the event date, along with two indicators of information control (government censorship and self-censorship), internet availability, and regime type. The purpose of M1 is to demonstrate that reporting delays are not uniform across countries but instead vary systematically with country-level characteristics. 

\begin{figure}[h]
    \centering
    \includegraphics[width=0.9\textwidth]{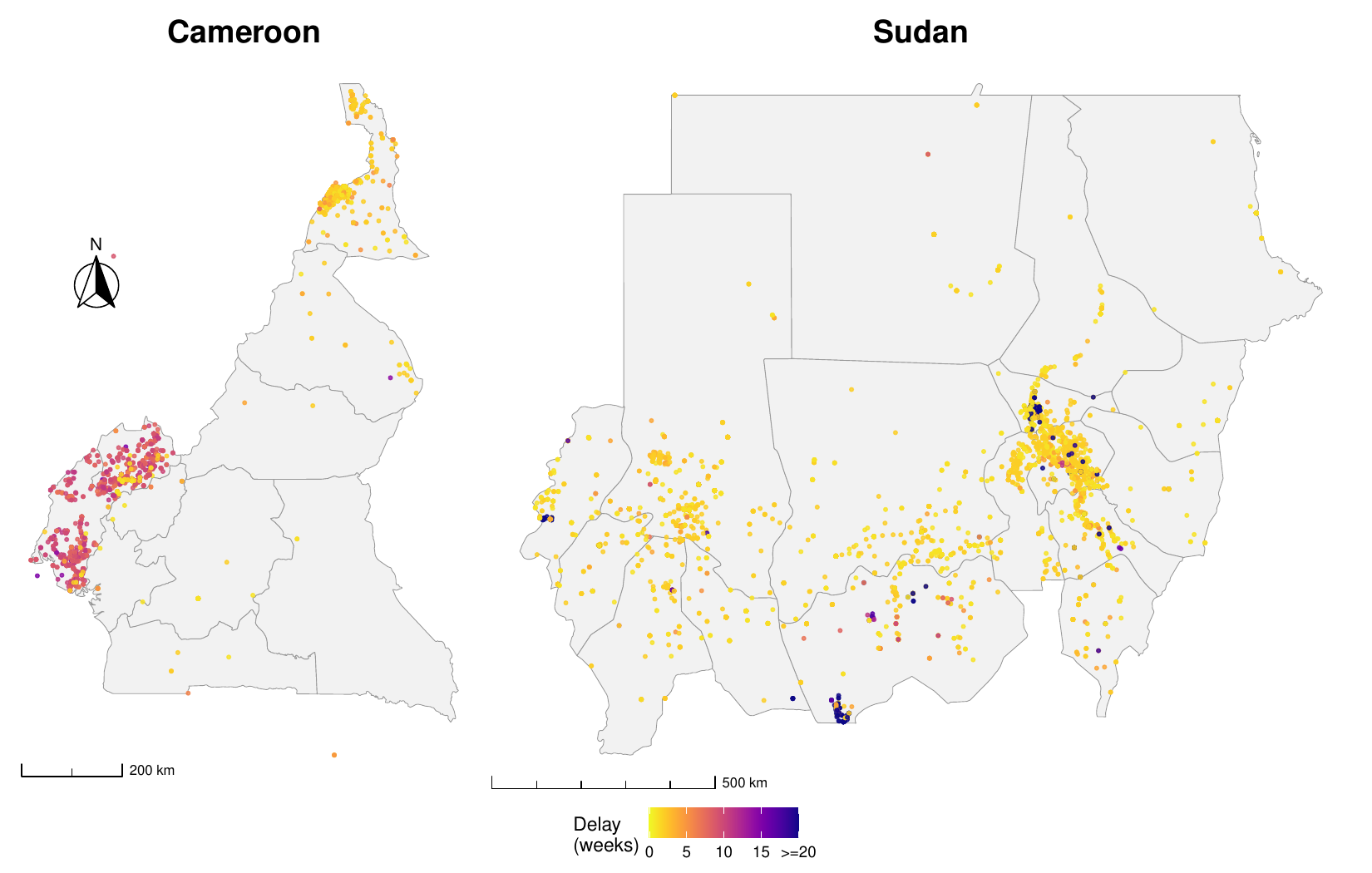} 
    \caption{Distribution of reporting delays in Cameroon and Sudan. Sudan shows a higher share of events reported early, whereas Cameroon has many more events with long reporting delays. Within-country variation is also visible: in Cameroon, delays are particularly pronounced in the South-West, while in South Sudan several highly delayed events appear near the southern border. These patterns highlight the presence of within-country heterogeneity in reporting delays.}
    \label{fig:cam_sud}
\end{figure}

In the second step, we develop the \textbf{event-level} model (M2), which incorporates a set of event-level covariates: the logarithm of the population within a 50 km radius of the event location, distance to the national capital, the number of events and fatalities during the week of occurrence, and the distance to the nearest international border. The goal of M2 is to evaluate how local characteristics shape reporting delays once broader country-level differences have been accounted for. To this end, we estimate M2 separately and exemplarily for two selected countries. This country-specific estimation serves two purposes. First, it allows us to assess the extent of within-country heterogeneity, demonstrating that reporting delays vary substantially between events even when national structural factors are kept constant. Second, comparing country-specific estimates highlights that the influence of event-level covariates is not uniform across contexts. Both the direction and magnitude of the effects differ between countries. %This finding reinforces the insights from Model M1, showing that variation in reporting delays locally differ across different countries. 
The motivation for adopting a country-specific modeling strategy is illustrated in Figure~\ref{fig:cam_sud}, which shows notable differences in the distribution of reporting delays between and within countries, particularly when comparing Sudan and Cameroon. Based on these patterns, we estimate model M2 for the countries shown in the figure and present the corresponding results in the next section. 

In both models, we use a proportional hazards framework \citep{cox1972regression}. Despite the assumption of proportional hazards, it provides a robust and interpretable approach for analyzing duration-type data. Because reporting delays are measured in discrete time units (weeks), we adopt a discrete-time specification based on the grouped proportional hazards model \citep{kalbfleisch1973marginal, prentice1978regression}. This formulation approximates the continuous-time Cox model and can be estimated using a generalized linear model (GLM) with a complementary log-log (cloglog) link function \citep{tutz2016modeling}. An outline of this approach in international relations and peace research is provided, for example, in \citet{beck1998taking}. Within this framework, each event with an observed delay \( D_i = d_i \) is represented by \( d_i \) observations, one for each week \( t = 1, 2, \ldots, d_i \). For each time point $t$, we define a binary response variable \( Y_{it} \), with \( Y_{it} = 0 \) for \( t < d_i \) (the event has not yet been reported) and \( Y_{it} = 1 \) at \( t = d_i \) (the event is reported). Events exceeding the censoring threshold (\( d_i > 20 \)) are right-censored at \( t = 20 \), with \( Y_{i 20} = 0 \). Thus, an event with a delay of \( D_i = 4 \) yields four observations: \( Y_{i1} = Y_{i2} = Y_{i3} = 0\), and \( Y_{i4} = 1 \). As a result, each \( Y_{it} \) follows a Bernoulli distribution with parameter \( \pi_{it} \), denoting the conditional probability that event \( i \) is reported in week \( t \), given that it has not been reported previously. 

All binary observations associated with event \( i \) share the same covariate vector \( X_i^{(k)} \), where \( k \in \{1,2\}\) indexes the model \( M_k \) specification and thus the employed covariates. In addition, we include a temporal variable \( t \) for each \( Y_{it} \) to capture the baseline hazard over weeks. We allow this temporal effect to interact with the event type in both models, reflecting evidence that reporting dynamics differ systematically between event categories. Accordingly, we estimate event-type–specific baseline hazards.

Rather than imposing a fixed parametric form, we estimate the baseline hazard non-parametrically by incorporating a smooth function within a Generalized Additive Model (GAM) framework \citep{wood2017generalized}. This approach provides greater flexibility in capturing the temporal dynamics of the reporting behavior by allowing the baseline hazard to vary smoothly over time. Smooth functions are also applied to the continuous covariates in both models, enabling potentially non-linear relationships to be modelled without requiring strong functional assumptions.

Formally, we assume that the reporting times of events are mutually independent and model the data as follows,
\begin{equation}
Y_{it} \sim \text{Bernoulli}(\pi_{it}),
\end{equation}
where \( \pi_{it}\) is the discrete-time hazard. It follows that \( \mathbb{E}(Y_{it}) = \pi_{it} \), i.e. \( \pi_{it}\) is the expected value of \( Y_{it} \) and represents the conditional probability.

The relationship between the hazard \( \pi_{it} \) and the covariates in \( X_i^{(k)} \) is defined via the complementary log-log (cloglog) link and modeled under specification \(M_k\) as
\begin{equation}
\label{eq:eq1}
\pi_{it} = 1 - \exp\left(-\exp(\eta^{(k)}_{it})\right).
\end{equation}
The additive predictor \( \eta^{(k)}_{it} \) depends on the set of covariates included in model \( M_k \).  

For the country-level model \( M_1 \), the predictor takes the form
\begin{equation}
\label{eq:m1}
\begin{aligned}
\eta^{(1)}_{it} = {} &
b^{(1)}(t, \text{type}_i) 
+ s^{(1)}_2(\text{logGDP}_i) 
+ s^{(1)}_3(\text{logPOP}_i) 
+ s^{(1)}_4(\text{logFATALITY}_i) \\
& + s^{(1)}_5(\text{GOVCENSOR}_i) 
+ s^{(1)}_6(\text{SELFCENSOR}_i) \\
& + \beta_{\text{internet}} \text{internet}_i 
+ \beta_{\text{regime}} \text{regime}_i .
\end{aligned}
\end{equation}
Here $b^{(1)}(t, \text{type}_i)$ represents an event-type–specific baseline temporal effect, defined as $b^{(1)}(t, \text{type}_i) = \beta_0^{(1)} + s^{(1)}_0(t, \text{type}_i)$, where $s^{(1)}_0$ is a factor–smooth interaction that estimates a separate smooth function of 
$t$ for each event type. The terms $s^{(1)}_2$, $s^{(1)}_3$, and $s^{(1)}_4$ are smooth functions of the logarithm of GDP per capita, total population, and cumulative fatalities associated with the relevant sub-event from January 1, 2024, to the event date, respectively. The functions $s^{(1)}_5$ and $s^{(1)}_6$ denote smooth effects of government censorship and self-censorship. The coefficients $\beta_{\text{internet}}$ and $\beta_{\text{regime}}$ represent linear effects of internet availability and regime type.

For the event-level model \( M_2 \) the addictive predictor is specified as
\begin{equation}
\label{eq:m2}
\begin{aligned}
\eta^{(2)}_{it} = {} &
b^{(2)}(t, \text{type}_i) 
+ s^{(2)}_2(\text{logBORDER}_i) 
+ s^{(2)}_3(\text{logPOP50km}_i) \\
&+ s^{(2)}_4(\text{logDISTANCE}_i)
+ s^{(2)}_5(\text{logFATALITY}_i) .
\end{aligned}
\end{equation}
Analogously, $b^{(2)}(t, \text{type}_i) = \beta_0^{(2)} + s^{(2)}_0(t, \text{type}_i)$ where $s^{(2)}_0$ is a factor–smooth interaction allowing the baseline hazard to vary flexibly over time for each event type. The smooth terms $s^{(2)}_2$, $s^{(2)}_3$, and $s^{(2)}_4$ capture the effects of the logarithm of distance to the nearest international border, population within a 50 km radius, and distance to the national capital, respectively. The term $s^{(2)}_5$ is a smooth function of the logarithm of fatalities during the week of occurrence.

Across both models, covariate effects are assumed to be time-invariant, i.e. all covariates \( X_i^{(k)} \) are treated as fixed characteristics of event \( i \) and their effects are assumed to be constant across time. This assumption simplifies interpretation and aligns with our goal of identifying the structural determinants of reporting delays. To ensure identifiability, all smooth terms are centered to integrate out to zero. Model estimation is carried out in \texttt{R} using the \texttt{mgcv} package \citep{wood2017generalized}, which provides efficient algorithms for GAM estimation and supports smooth temporal effects.

\section{Results}

\subsection{Across country effect}

\begin{figure}[ht]
    \centering
    \includegraphics[width=0.9\textwidth]{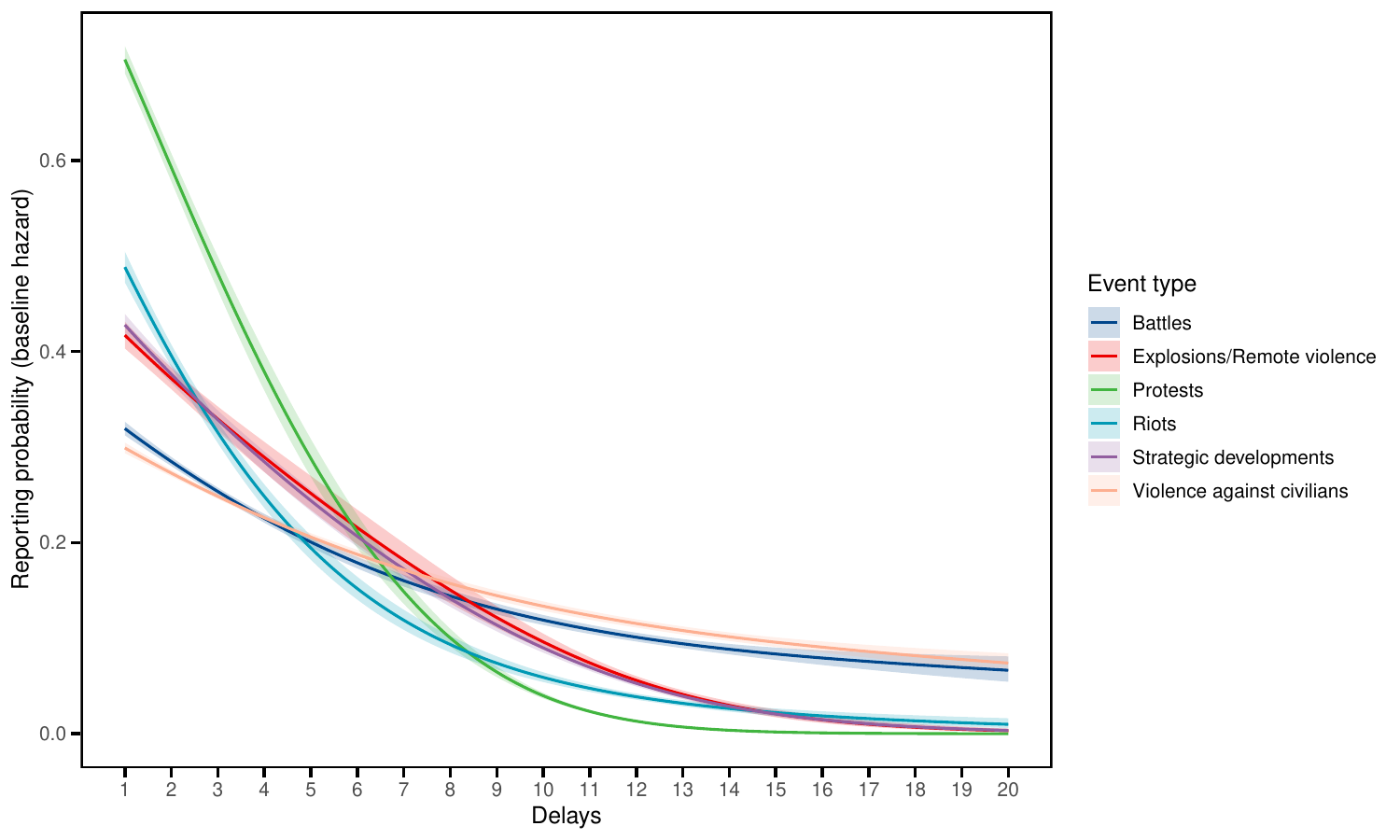} 
    \caption{Event-type–specific baseline hazard derived from Model M1}
    \label{fig:base}
\end{figure}

To illustrate the reporting dynamics captured by Model M1, we first examine the event-type–specific baseline hazard $b^{(1)}(t, \text{type}_i)$. Figure~\ref{fig:base} displays the corresponding baseline hazards, with all other covariate effects set to zero for comparability. The curves are shown on the probability scale and represent the estimated probability that an event of a given type is reported in week $t$, conditional on not having been reported earlier.

Across all types of events, the baseline hazard declines over time, indicating that events are more likely to be reported shortly after their occurrence than after long delays. However, both the shape and the steepness of this decline vary substantially between event types. Protests exhibit the highest probability of early reporting and thus the lowest probability of long reporting delays. This pattern is consistent with the high visibility of protests, as protest organizers often actively seek media attention, resulting in comparatively faster reporting. 

By contrast, battles and violence against civilians exhibit the lowest probabilities of early reporting and the highest probabilities of long delays, with comparatively gradual declines in their baseline hazards. Riots, strategic developments, and explosions/remote violence show steeper declines and comparatively low probabilities of long reporting delays. Together, these patterns highlight pronounced heterogeneity in baseline reporting dynamics across event types. Once country-level covariates are incorporated in Model M1, this heterogeneity becomes even more pronounced, underscoring the extent to which national structural characteristics further shape reporting dynamics.

\begin{table}[ht]
\centering
\begin{threeparttable}
\begin{tabularx}{0.9\textwidth}{@{}Xcc@{}}
\toprule
Term & Estimate & SE \\
\midrule
\multicolumn{3}{l}{\textbf{Regime}} \\
\quad Electoral Autocracy & -0.207$^{***}$ & 0.014 \\
\quad Democracy & 0.173$^{***}$ & 0.035 \\
\addlinespace
\multicolumn{3}{l}{\textbf{Internet}} \\
\quad Yes & 0.048$^{*}$ & 0.020 \\
\bottomrule
\end{tabularx}
\begin{tablenotes}
\small
\item \textit{Note:} Significance codes: $^{***} p < 0.001$, $^{**} p < 0.01$, $^{*} p < 0.05$
\end{tablenotes}
\caption{Estimated effects of categorical covariates of Model M1. The reference category for Regime is “Closed Autocracy,” and for Internet access is “No.”}
\label{tab:param_country}
\end{threeparttable}
\end{table}

Table~\ref{tab:param_country} reports the estimated effects of regime type and internet availability from Model M1. Using closed autocracy as the reference category, we observe that democracy has a significant positive effect on reporting, indicating that events tend to be reported faster in democratic political environments. By contrast, the effect for electoral autocracy is significantly negative. As expected, internet availability is also positively associated with reporting timeliness: events are more likely to be reported earlier in contexts with reliable internet access, as reflected by the significantly positive coefficient relative to the no-internet reference category.

\begin{figure}[ht]
    \centering
    \includegraphics[width=0.9\textwidth]{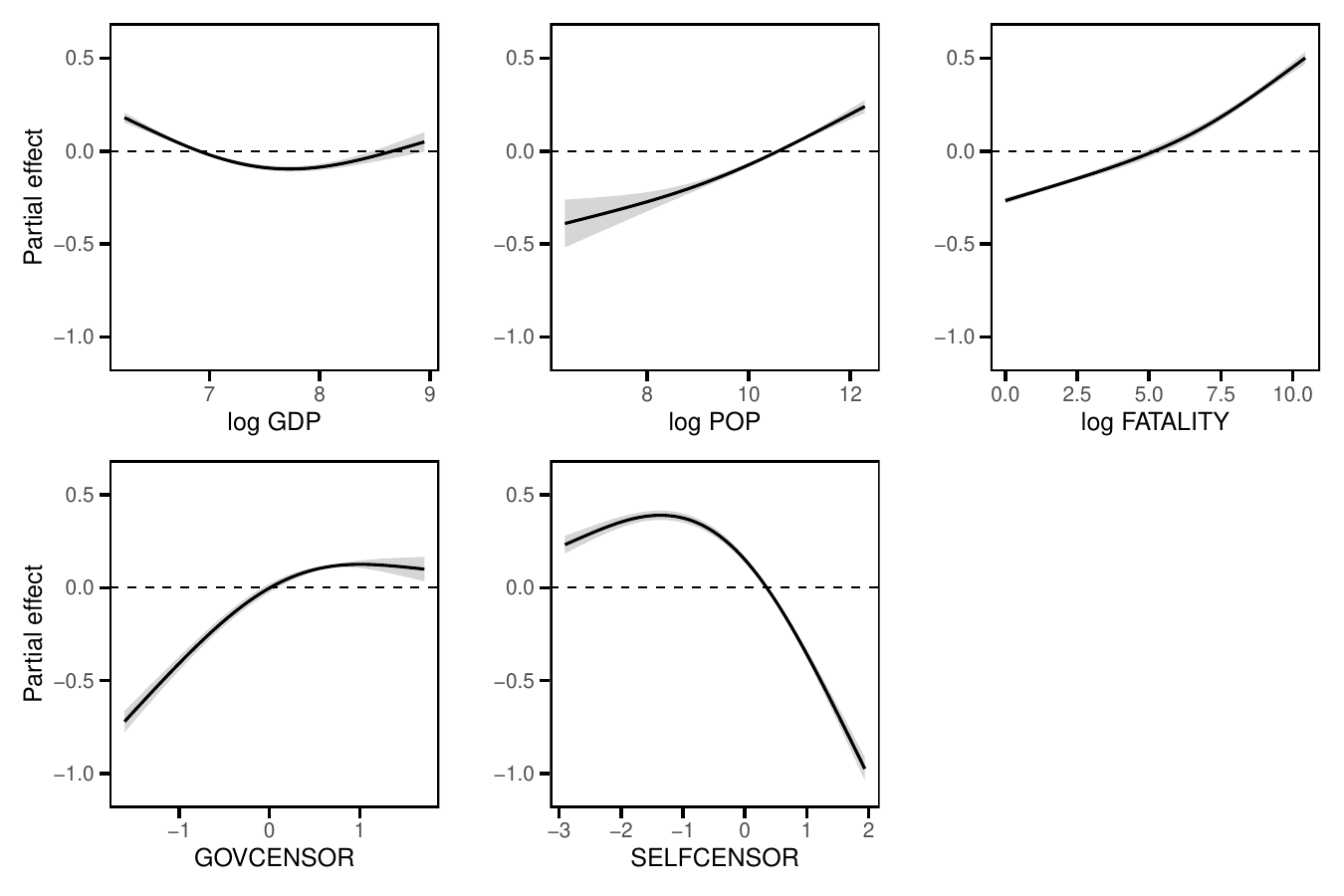} 
    \caption{Smooth covariate effects}
    \label{fig:country_smooth}
\end{figure}

In addition to these categorical covariates, Figure~\ref{fig:country_smooth} displays the estimated partial effects of the smooth terms included in Model M1. We begin with the effect of log GDP per capita. The estimated smooth is nearly flat across its range, indicating that -- conditional on the other covariates in the model -- the partial effect of economic development on reporting delays is limited. This suggests that once factors such as regime type and information infrastructure are accounted for, GDP per capita explains little additional variation in reporting timeliness.

By contrast, the smooth effects of log population, cumulative fatalities, and government censorship exhibit increasing trends. These patterns indicate that higher population size, higher accumulated fatalities within a sub-event, and lower levels of government censorship are associated with a higher probability of earlier reporting. This finding is consistent with the idea that events occurring in more populous contexts or involving higher levels of violence tend to be more visible and therefore more likely to be reported promptly.

Media self-censorship, however, shows a decreasing partial effect, implying that lower levels of self-censorship are associated with longer reporting delays. One possible interpretation is that in contexts characterized by higher self-censorship, information may be suppressed entirely rather than entering the reporting process with a delay. In more permissive media environments, on the contrary, reporting can be delayed as information is subject to verification, editorial processes, or competing news priorities. We emphasize that these interpretations are suggestive and reflect associations captured by the model rather than causal effects.

\subsection{Within country effect}

We now turn to the country-specific model M2, fitted separately for Cameroon and Sudan, to examine within country heterogeneity in reporting delays. Comparing these two cases also highlights cross-country divergence in reporting dynamics, confirming the heterogeneity between countries discussed in the previous subsection. In other words, heterogeneity in reporting dynamics is present both between and within countries. 

\begin{figure}[ht]
    \centering
    \includegraphics[width=0.8\textwidth]{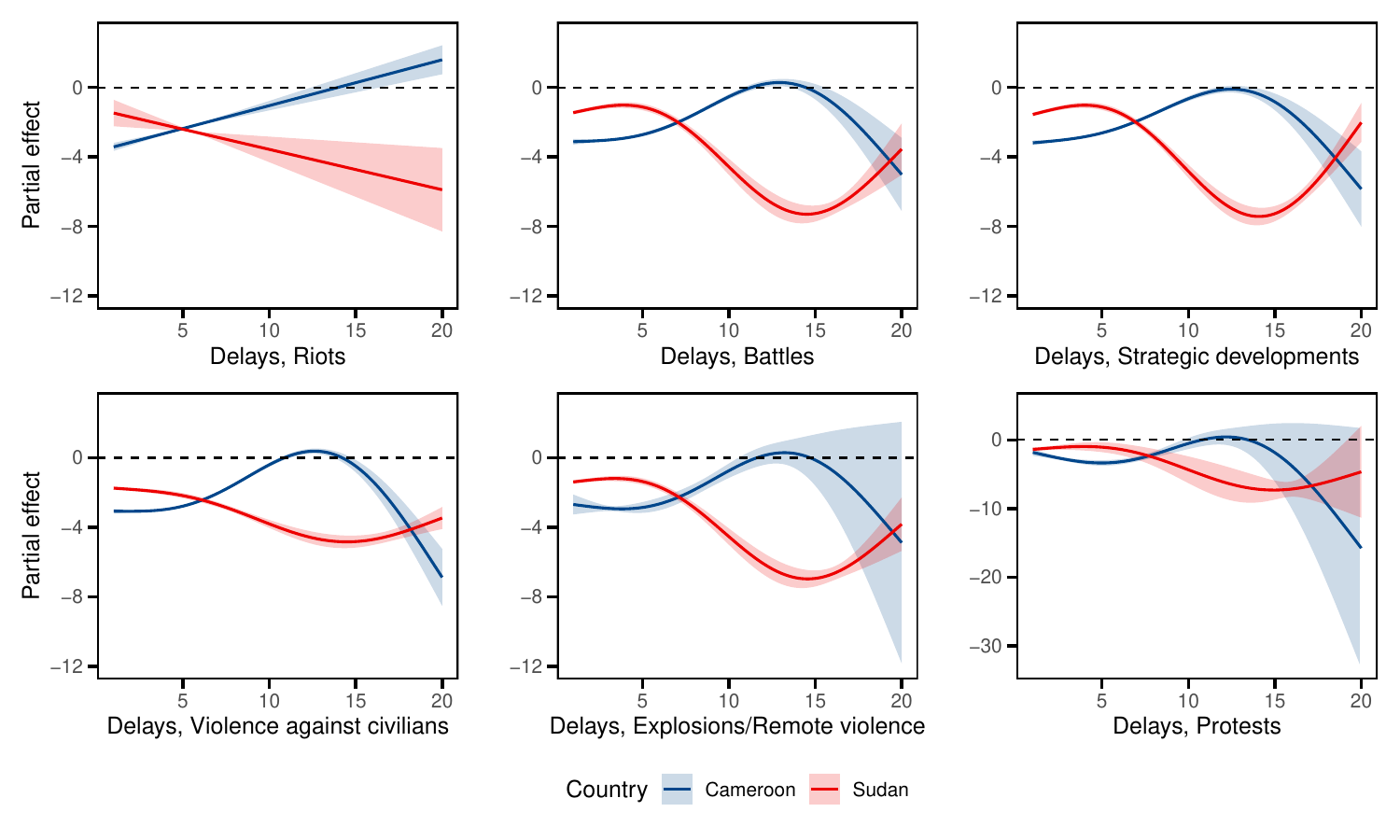} 
    \caption{Country–specific baseline hazard derived from Model M2}
    \label{fig:baseline}
\end{figure}

As in the previous subsection, we begin by examining the event-type–specific baseline hazard for the two country-specific versions of Model M2.
Figure~\ref{fig:baseline} displays these baseline hazards, with event types now shown in separate plots, to allow for a comparison between Cameroon and Sudan. To be specific, the plots show $b^{(2)}(t, \text{type}_i)$ for each type of event. The baseline hazards differ sharply between Sudan and Cameroon between event types, revealing distinct dynamics in reporting probabilities over time.
%With the exception of riots, the overall shapes are broadly similar across event categories within each country. 
The most striking difference occurs for riots. In Sudan, the baseline hazard decreases over time, indicating that riots are more likely to be reported shortly after occurrence, whereas in Cameroon the hazard increases, suggesting a higher likelihood of delayed reporting. For the remaining event types, Sudan broadly exhibits U-shaped baseline patterns, whereas Cameroon displays reverse U-shaped patterns. In Sudan, reporting is concentrated in the first five weeks, declines until roughly week 15, and then increases again, suggesting the coexistence of timely reporting and very long delays, with fewer events reported at intermediate durations. In contrast, Cameroon shows relatively few events reported in the earliest weeks, with reporting probabilities peaking at intermediate delays around week 13 before declining thereafter. This pattern suggests that events in Cameroon are most often reported with medium-length delays rather than very short or very long delays.

\begin{figure}[ht]
    \centering
    \includegraphics[width=0.7\textwidth]{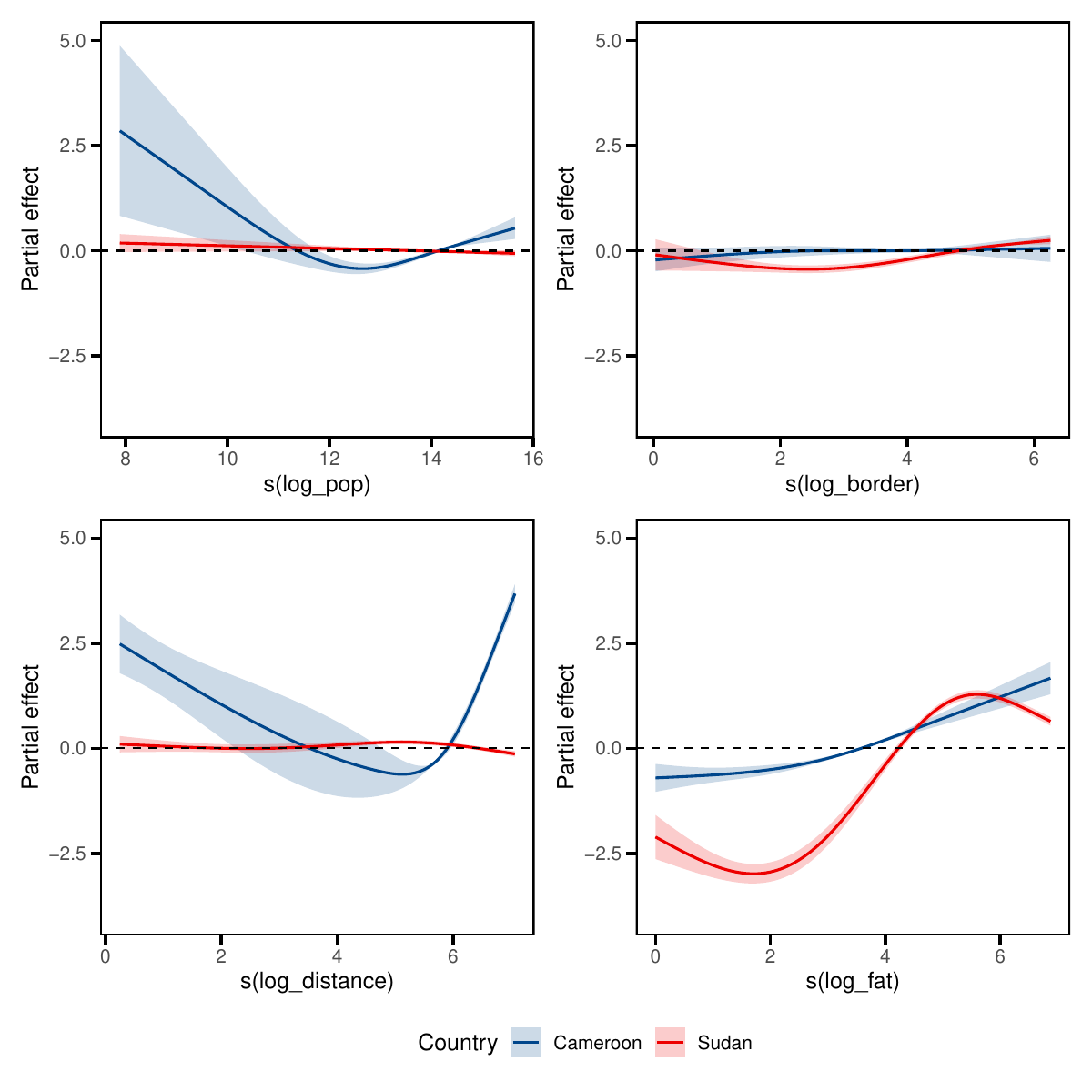} 
    \caption{Smooth terms}
    \label{fig:base_event}
\end{figure}

Figure~\ref{fig:base_event} illustrates the estimated smooth effects of the covariates included in Model~M2 on reporting delays. Overall, Cameroon exhibits substantially greater heterogeneity than Sudan in the effects of local population size (within a 50 km radius) and distance to the national capital. In Sudan, the estimated effects of these covariates are close to zero across their entire ranges.

In Cameroon, reporting is most timely in areas that are either sparsely populated or highly populated, whereas locations with intermediate population density display are associated with slower reporting. By contrast, population size has a negligible effect on reporting delays in Sudan. Distance to the nearest international border has little influence on reporting timeliness in both countries. Distance to the capital, however, shows a pronounced nonlinear effect in Cameroon but remains effectively flat in Sudan. In Cameroon, this relationship follows a U-shaped pattern: events occurring near the capital tend to be reported more quickly, reporting slows down as distance increases, reaching a minimum around log(distance) = 5.5, and then accelerates again for events located farther from the capital. This pattern suggests that both central and remote locations benefit from faster reporting, whereas intermediate regions experience longer delays.

Beyond spatial characteristics, Model~M2 also captures the effects of event severity through fatality counts. Unlike the spatial covariates, fatalities exhibit substantial variation in both countries. In general, higher death counts are associated with faster reporting, consistent with greater visibility of more severe events. However, this relationship is more strongly nonlinear in Sudan, indicating that the effect of fatalities on reporting timeliness varies between contexts.

Together, these results reveal substantial heterogeneity within the country in reporting delays. Importantly, both the magnitude and the functional form of covariate effects differ across countries, underscoring the need for country-specific modeling approaches when analyzing and accounting for reporting dynamics in conflict event data.

\section{Discussion \& Conclusion}
Our analysis demonstrates that ACLED data are subject to systematic reporting delays. While such delays are to  be expected in any near-real-time event database that issues periodic releases -- for example weekly updates -- due to an inherent trade-off between timeliness and complete event capture, our intention was to show that these delays are neither random nor uniform. Instead, delays exhibit substantial heterogeneity and vary systematically between national and event-level contexts. Crucially, the delay process is highly structured and therefore amenable to statistical analysis.

A central finding of this study is that delays depend on both country-level and event-level factors. At the country level, socio-economic conditions, regime type, information accessibility, and media-related constraints shape the reporting environment and influence how quickly events enter the data set. At the event level, attributes such as event type, fatalities, population exposure, and spatial location affect reporting timeliness. 
%A central finding of this study is that reporting delays depend on both country-level and event-level factors. At the country level, socio-economic factors, political regime type, information accessibility, and media-related constraints shape the overall reporting environment and influence how quickly events enter the dataset. At the event level, characteristics such as event type, fatalities, population exposure, and spatial location affect reporting timeliness. 
Importantly, these event-level effects are not stable across contexts. Our selected country-specific analyses indicate that the direction, magnitude, and functional form of event-level covariate effects vary substantially between countries. Our findings underscore that reporting delays are generated by context-dependent processes rather than a single homogeneous mechanism, thus highlighting the limitations of a single global delay model.

We acknowledge several limitations. First, our contribution is descriptive rather than corrective. We characterize the distribution and determinants of delays, but we do not offer a statistical "cure". However, we argue that a clear description of the delay structure is a necessary prerequisite to corrective procedures such as nowcasting. Second, while we document pronounced within-country heterogeneity using Cameroon and Sudan as illustrative cases, we do not estimate event-level models for every African country.
%We do not propose or implement a statistical “cure” for delayed reporting, but instead focus on characterizing the distribution and determinants of reporting delays. While this is a limitation, we emphasize that understanding the structure of delayed reporting is required first, before deriving correction tools like nowcasting. Second, while our analysis demonstrates pronounced within-country heterogeneity using two illustrative cases (Cameroon and Sudan), we do not estimate country-specific event-level models for all African countries. 
Our selection of these two cases is intended to establish the existence and relevance of context-dependent delay mechanisms, not to exhaustively map the respective heterogeneities across the continent. Third, although our covariate set is theoretically motivated and inspired by previous contributions on reporting biases,  additional factors such as source diversity, media competition, linguistic barriers, or international attention could also influence delays. We intend to consider such factors more exhaustively in follow-up studies.
%Third, although we include a theoretically motivated set of covariates, other factors such as source diversity, media competition, linguistic barriers, or international attention may also play an important role in shaping reporting delays and are not considered here. 
Given our primary objective of demonstrating that delays are systematic and heterogeneous, the exclusion of additional covariates does not undermine our core findings. 

The results of this paper provide a foundation for future methodological work to correct for reporting delays. A natural next step is to develop nowcasting approaches for conflict event data that explicitly account for delayed reporting by adjusting observed counts to include events that are expected but not yet reported. As demonstrated in \cite{schneble2021nowcasting} or \cite{fritz2024statistical}, such adjustments can be implemented numerically using available tools. However, our findings suggest that this is a nontrivial task. Because delay distributions differ across countries and depend on multiple interacting factors, effective nowcasting models must accommodate substantial heterogeneity.
%A natural next step is the development of nowcasting approaches for conflict event data that explicitly account for delayed reporting. Such methods would adjust observed event counts by incorporating information about the expected number of events yet to be reported. In \cite{schneble2021nowcasting} or \cite{fritz2024statistical}, we show how this can be incorporated numerically in available tools. However, our findings indicate that this task is nontrivial: because delay distributions vary across countries and depend on multiple interacting factors, effective nowcasting models must accommodate substantial heterogeneity. 
Future research should therefore explore country-specific delay models in more detail and investigate strategies for combining them into a unified framework suitable for regional or continental nowcasting. Identifying additional predictors of delay and developing rigorous methods to pool information across contexts will be key challenges in this endeavor.

%In sum, this paper shows that reporting delays in conflict data are structured, heterogeneous, and shaped by identifiable contextual factors. Recognizing and modeling these delays is a necessary first step toward improving real-time conflict monitoring and forecasting. Addressing them through appropriate nowcasting techniques represents an important and promising direction for future research.

%HETERO COUNTRY THEN HETERO IND THEN NEED FOR COUNTRY SPECIFIC MODEL
%WE DID NOT ACCOUNT FOR DIFFUSION/SELF EXCITING PROP OF CONFLICT REPORTING
\section*{Acknowledgments}

This work is supported by the German Academic Exchange Service (DAAD) under the research grant for doctoral studies.

\bibliographystyle{apacite}
\bibliography{references} 

\clearpage

\begin{titlepage}
    \centering
    \vspace*{4cm} % space from top
    {\LARGE Appendix for Assessing Reporting Delays in ACLED Conflict Event Data \par}
    \vspace{1cm} % space below title (optional for aesthetics)
\end{titlepage}

\appendix
This appendix accompanies the main paper \textit{"Assessing Reporting Delays in ACLED Conflict Event Data"}. It provides three principal appendices. First, we describe how reporting delays are calculated. Secondly, we explain the data cleaning. And finally, we present basic exploratory data analyses.

\section{Delay calculation and data structure}

The time variables used to calculate reporting delays are as follows:

\begin{itemize}
    \item \texttt{event\_date}: The date on which the conflict event occurred.
    \item \texttt{timestamp\_date}: The date on which the event was updated in the database. This may correspond either to the first report of the event or to a later update of an already reported event.
    \item \texttt{date\_id}: The date on which the dataset was downloaded.
\end{itemize}

The analysis focuses exclusively on newly recorded events. These are identified by comparing the weekly updates with previously uploaded data version, going back to the initial dataset downloaded on June 30, 2024. Events that appear in later downloads but were absent in earlier ones are classified as new. If an event initially identified as new appears in later downloads with a more recent \textit{timestamp\_date}, indicating an update to the previously recorded event, we retain the earlier \textit{timestamp\_date} to capture its first appearance in the database. For these newly reported events, the reporting delay is calculated as the difference in weeks between the \textit{timestamp\_date} and the \textit{event\_date}.

The list of countries included in the analysis is determined based on event frequency. Specifically, countries with fewer than 10 events that occurred in 2023 are excluded. We choose 2023 because we assume that events from that year are fully reported. Under the assumption of stationarity in the number of conflict events and independence of the delay distribution from the reporting time, the number of events that occurred in 2023 provides a reasonable estimate of the total number of events -- whether reported on time or with delay -- that can be expected in a typical year. Countries with very few events in 2023 are therefore expected to have few reported events in our delay dataset as well.  

The delay dataset itself consists of events that were reported between 30 June 2024 and 01 June 2025. These events may have occurred at any point in time, as reporting delays mean that older events can appear in later dataset releases. Figure~\ref{fig:figure_2023} compares the number of events that occurred in 2023 with the number of events in the delay dataset. Based on the 2023 counts, we omit countries with fewer than 10 events in that year. We also omit Mayotte, La Réunion, and the British Indian Ocean Territory, as these are overseas territories and not considered independent African countries. Table~\ref{tab:tab1} presents the number of events per country and indicates which countries have been excluded from the analysis.

\begin{table}[h]
\caption{Countries by event counts occurred in 2023}
\label{tab:tab1}
\centering

\begin{minipage}{0.48\textwidth}
\centering
\begin{tabular}{|l|r|}
\hline
Country & Count \\
\hline

Equatorial Guinea* & 2\\
Seychelles* & 4\\
Eritrea* & 6\\
Sao Tome and Principe* & 9\\
Gambia & 10\\
Republic of Congo & 12\\
Djibouti & 15\\
Lesotho & 18\\
Rwanda & 22\\
Botswana & 25\\
Guinea-Bissau & 33\\
Togo & 38\\
Comoros & 39\\
Cape Verde & 52\\
Tanzania & 65\\
Zambia & 66\\
Liberia & 67\\
Mayotte* & 73\\
Sierra Leone & 73\\
eSwatini & 84\\
Gabon & 87\\
Namibia & 103\\
Ivory Coast & 105\\
Mauritius & 117\\
Reunion* & 178\\
Chad & 182\\
Libya & 201\\
Malawi & 204\\

\hline
\end{tabular}
\end{minipage}
\hfill
\begin{minipage}{0.48\textwidth}
\centering
\begin{tabular}{|l|r|}
\hline
Country & Count \\
\hline

Egypt & 207\\
Angola & 217\\
Zimbabwe & 234\\
Mauritania & 251\\
Senegal & 289\\
Guinea & 312\\
Burundi & 334\\
Ghana & 368\\
Benin & 375\\
Algeria & 436\\
Central African Republic & 446\\
Mozambique & 452\\
Uganda & 568\\
Tunisia & 772\\
Niger & 795\\
Madagascar & 806\\
South Sudan & 1738\\
Ethiopia & 1767\\
South Africa & 2088\\
Morocco & 2214\\
Mali & 2248\\
Burkina Faso & 2359\\
Kenya & 2549\\
Democratic Republic of Congo & 3233\\
Cameroon & 3733\\
Somalia & 3794\\
Nigeria & 4713\\
Sudan & 6339\\

\hline
\end{tabular}
\end{minipage}
\begin{tablenotes}
\small
\item \textit{Note:} $^{*} Excluded$
\end{tablenotes}
\end{table}

\section{Data cleaning}

\begin{figure}[h]
  \centering
  \includegraphics[width=0.8\linewidth]{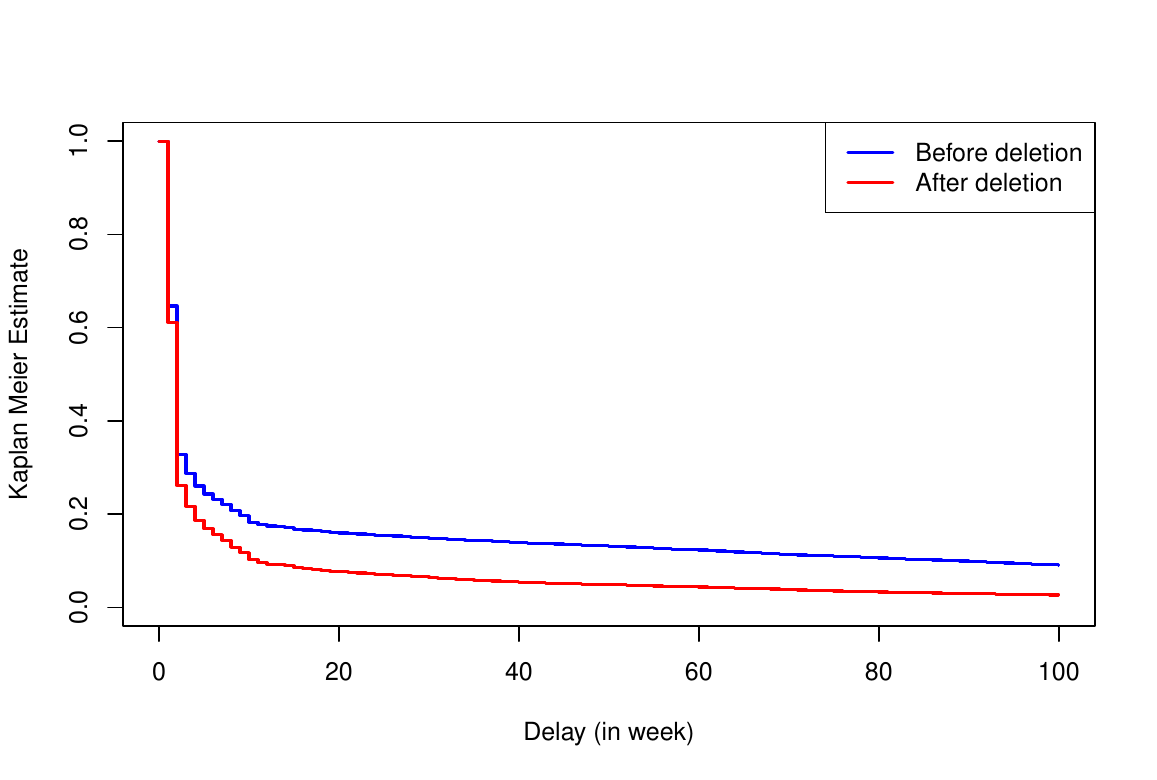}
  \caption{Kaplan Meier curve before and after deletion of historical events.}
  \label{fig:km_histo}
\end{figure}

\begin{figure}[h]
  \centering
  \includegraphics[width=0.8\linewidth]{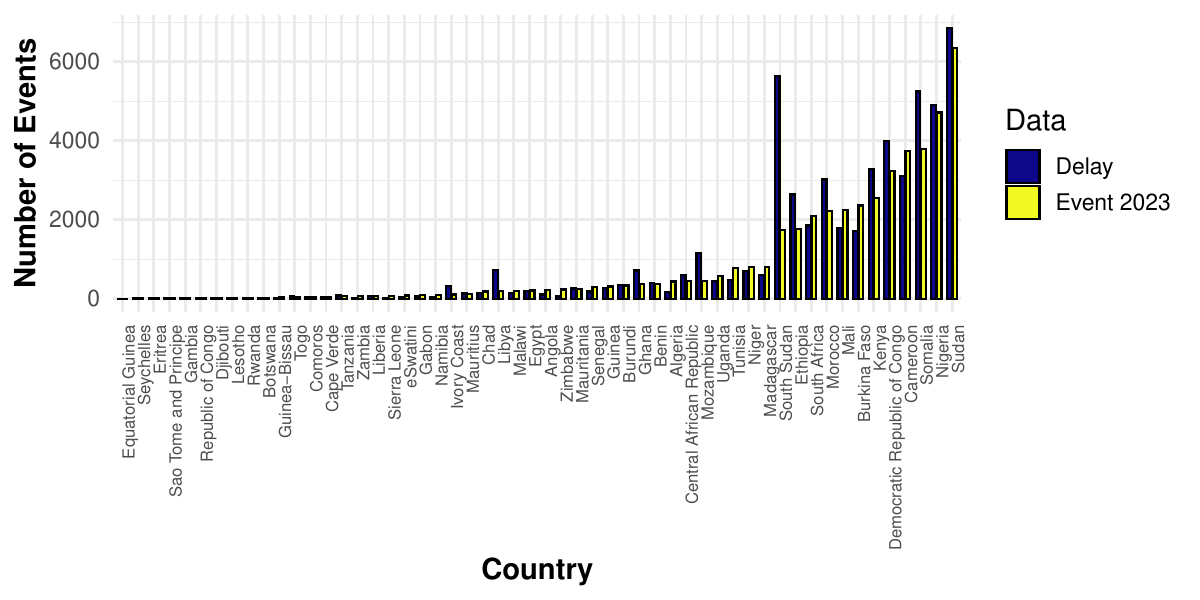}
  \caption{Number of events reported between 31 June, 2024 (blue) and number of events having occurred in 2023 (yellow).}
  \label{fig:figure_2023}
\end{figure}

Since the objective of the main paper is to analyze reporting delays in order to improve real-time and near-present monitoring, we treat historical event reports with particular care. Table~\ref{tab:histo_table} shows the number of reports delayed by more than 40 weeks on each reporting day. On a few specific days, we observe large batches of very delayed reports. Table~C lists these cases by country and source. It is clear that many of these are historical backlogs, as they originate from the same country and the same source on the same day. We remove those events for which we can be confident they represent historical uploads -- specifically, the large batches aggregated in the first two rows of Table~\ref{tab:histo_table}. The massive historical reports explain the wide discrepancy between the number of events reported between 30 June 2024 and 01 June 2025 for South Sudan and Somalia and the number of events that occurred in 2023 (Figure ~\ref{fig:figure_2023}). To assess the impact of this filtering, we compare Kaplan–Meier curves with and without these events in Figure~\ref{fig:km_histo}.

\begin{table}

\caption{Events with delays exceeding 40 weeks by country and source}
\label{tab:histo_table}
\centering
\begin{tabular}[t]{lllr}
\toprule
Timestamp\_date & Country & Source & Number\\
\midrule
2025-01-28 & South Sudan & Risk and Strategic Management, Corporation & 3901\\
2025-05-12 & Democratic Republic of Congo & Forum de Paix de Beni & 885\\
2025-02-04 & Somalia & Undisclosed Source & 787\\
2024-09-23 & Somalia & Undisclosed Source & 422\\
2024-09-24 & Somalia & Undisclosed Source & 312\\
2025-01-28 & Sudan & Risk and Strategic Management, Corporation & 119\\
\bottomrule
\end{tabular}
\end{table}

Before fitting smooth terms in the model, we examined the distributions of \(\log(1+\text{population}_{50\text{km}})\), \(\log(1+\text{distance to capital})\) and \(\log(\text{GDP per capita})\). 
Kernel density plots with individual observations marked along the axis reveal that all variables contain a small number of extreme values far from the bulk of the data. When such values are isolated, they can pull a smoother toward the edges of the scale, leading to less reliable estimates in the central range where most data lie.

To reduce the influence of these extreme points while keeping all observations, we winsorize each variable at thresholds chosen to match the nature of their extremes: for distance to the capital, values above the 99th percentile are set to the 99th percentile; for population within 50 km, values below the 1st percentile are set to the 1st percentile; and for GDP per capita, values below the 5th percentile are set to the 5th percentile. This preserves the main shape of each distribution while preventing a few isolated points from distorting the fitted curves.

\begin{figure}[h]
    \centering
    \begin{subfigure}[b]{0.45\textwidth}
        \centering
        \includegraphics[width=\textwidth]{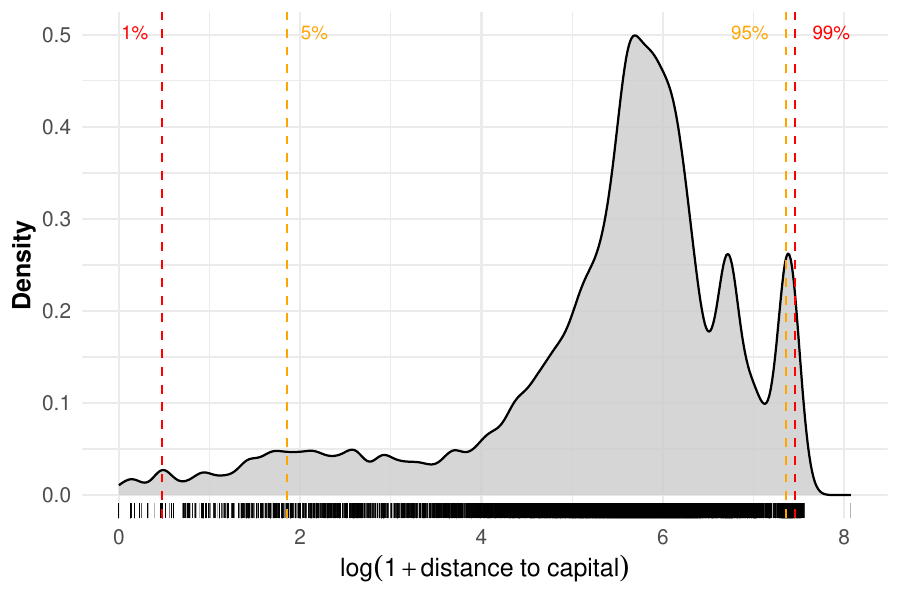}
        \caption{Log(1+distance to capital) before winsorizing}
        \label{fig:sub1}
    \end{subfigure}
    \hfill
    \begin{subfigure}[b]{0.45\textwidth}
        \centering
        \includegraphics[width=\textwidth]{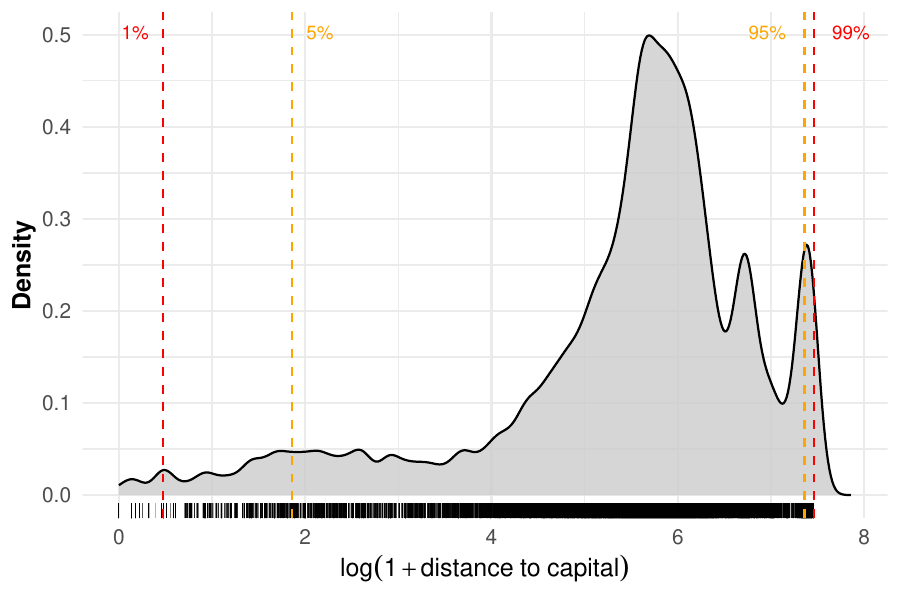}
        \caption{Log(1+distance to capital) after winsorizing}
        \label{fig:sub2}
    \end{subfigure}
    \hfill
    \begin{subfigure}[b]{0.45\textwidth}
        \centering
        \includegraphics[width=\textwidth]{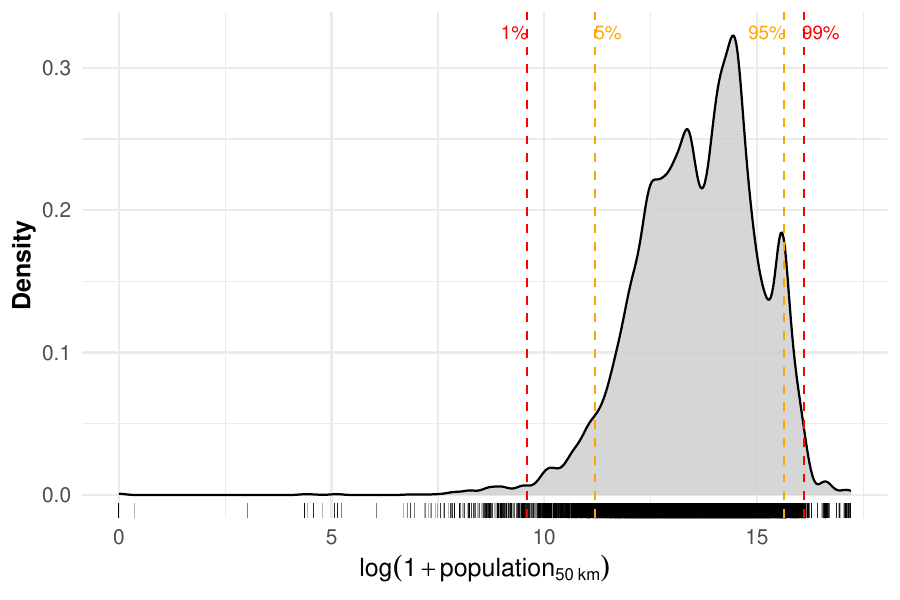}
        \caption{Log(1+population within 50km) before winsorizing}
        \label{fig:sub3}
    \end{subfigure}
    \hfill
    \begin{subfigure}[b]{0.45\textwidth}
        \centering
        \includegraphics[width=\textwidth]{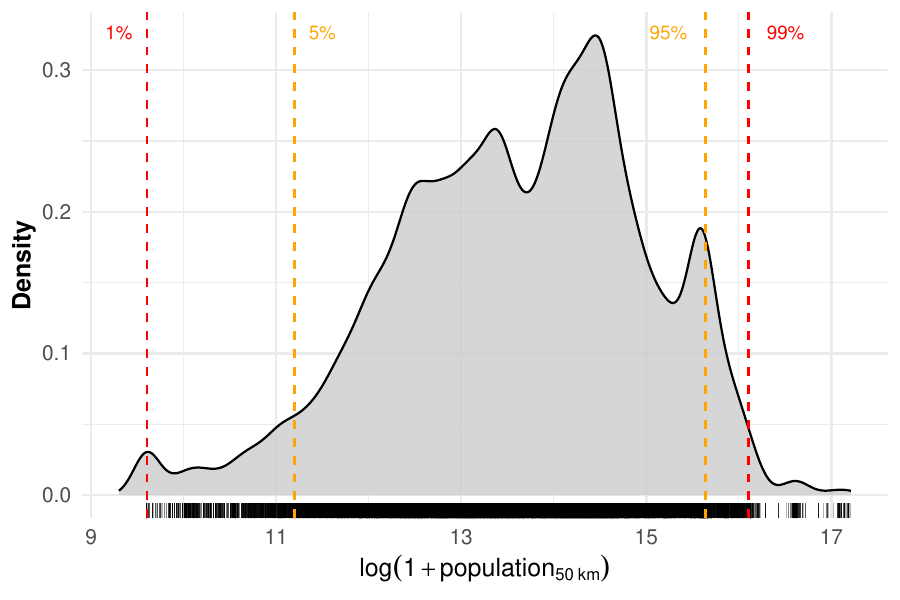}
        \caption{Log(1+population within 50km) after winsorizing}
        \label{fig:sub3}
    \end{subfigure}
    \hfill
    \begin{subfigure}[b]{0.45\textwidth}
        \centering
        \includegraphics[width=\textwidth]{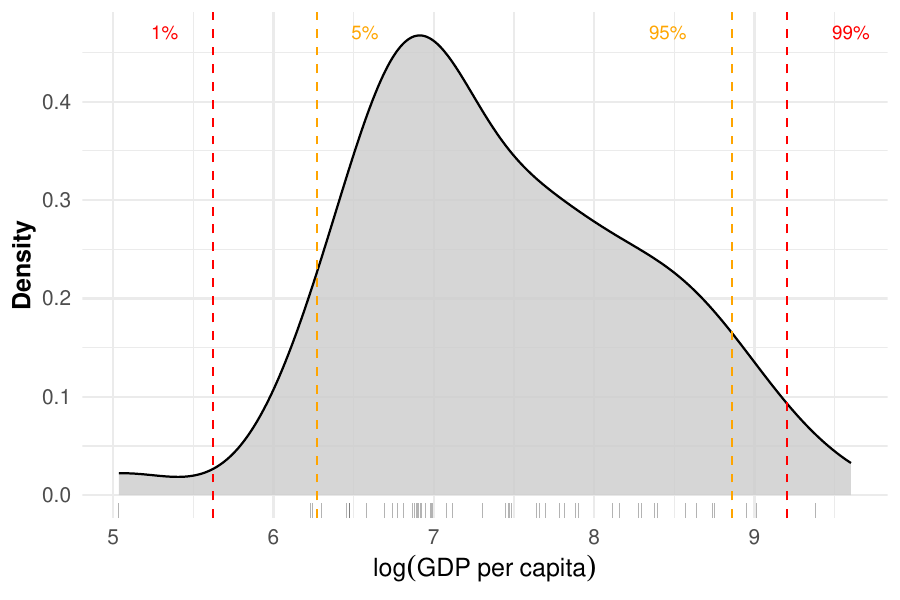}
        \caption{Log(GDP per capita) before winsorizing}
        \label{fig:sub3}
    \end{subfigure}
    \hfill
    \begin{subfigure}[b]{0.45\textwidth}
        \centering
        \includegraphics[width=\textwidth]{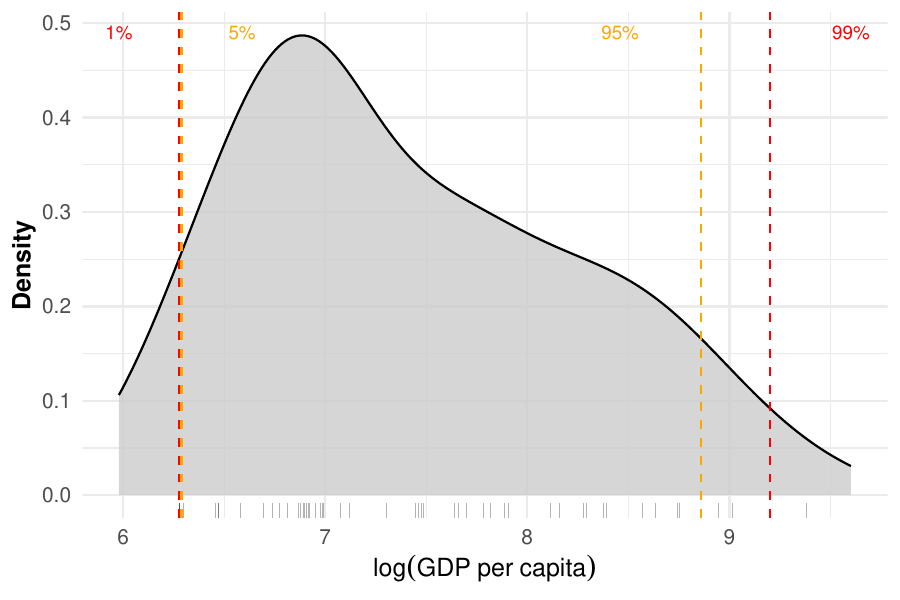}
        \caption{Log(GDP per capita) after winsorizing}
        \label{fig:sub3}
    \end{subfigure}
    \caption{Kernel density plots of \(\log(1+\text{population}_{50\text{km}})\), \(\log(1+\text{distance to capital})\), and \(\log(\text{GDP per capita})\).}
    \label{fig:winsorize}
\end{figure}

Figure~\ref{fig:winsorize} shows the estimated kernel densities before and after winsorizing. The overall patterns remain, but the extreme tails are shortened, allowing the smoother to focus on the range where most observations occur.

\section{Exploratory data analysis}

\begin{table}[h]
\centering
\caption{Summary Statistics}
\begin{tabular}[t]{lccccccc}
\toprule
Variable & N & Min & Q1 & Median & Mean & Q3 & Max \\
\midrule
Fatalities & 52793 & 0 & 0 & 0 & 1.402 & 1 & 370 \\
Population & 52793 & 0 & 311743 & 917897 & 1783390 & 2153886 & 29832576 \\
Distance & 52793  & 0 & 137.3 & 301.2 & 435.8 & 530.9 & 3213.8 \\
GDP & 50 & 153.9 & 965.2 & 1360.2 & 2428.0 & 3181.3 & 11871.7 \\
\bottomrule
\end{tabular}
\label{tab:summary_stats}
\end{table}

\begin{figure}[h]
  \centering
  \includegraphics[width=0.8\linewidth]{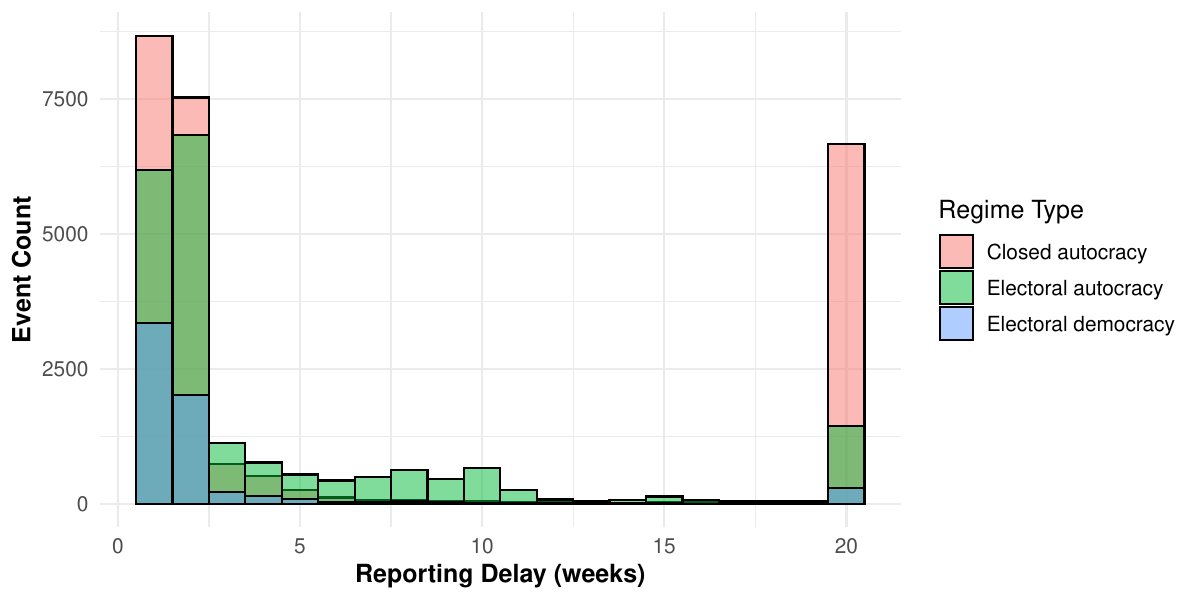}
  \caption{Reporting delay distribution by regime type.}
  \label{fig:regime}
\end{figure}

\begin{figure}[h]
    \centering
    \begin{subfigure}[b]{0.42\textwidth}
        \centering
        \includegraphics[width=\textwidth]{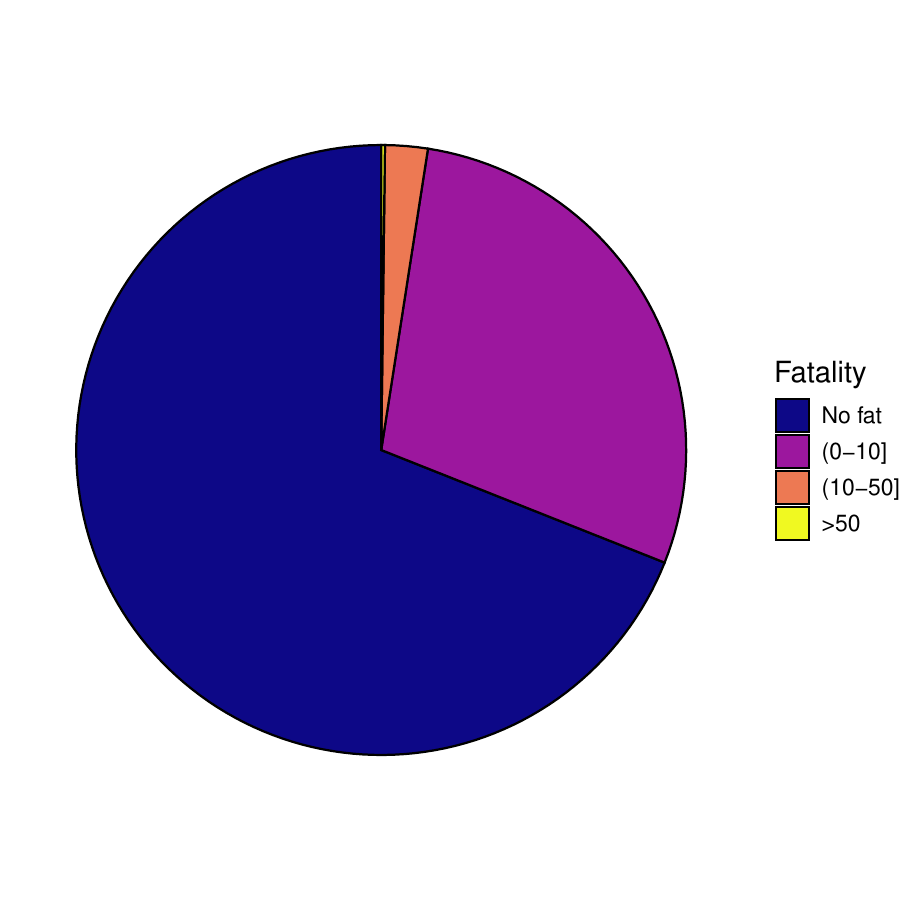}
        \caption{Fatality}
        \label{fig:fat}
    \end{subfigure}
    \hfill
    \begin{subfigure}[b]{0.45\textwidth}
        \centering
        \includegraphics[width=\textwidth]{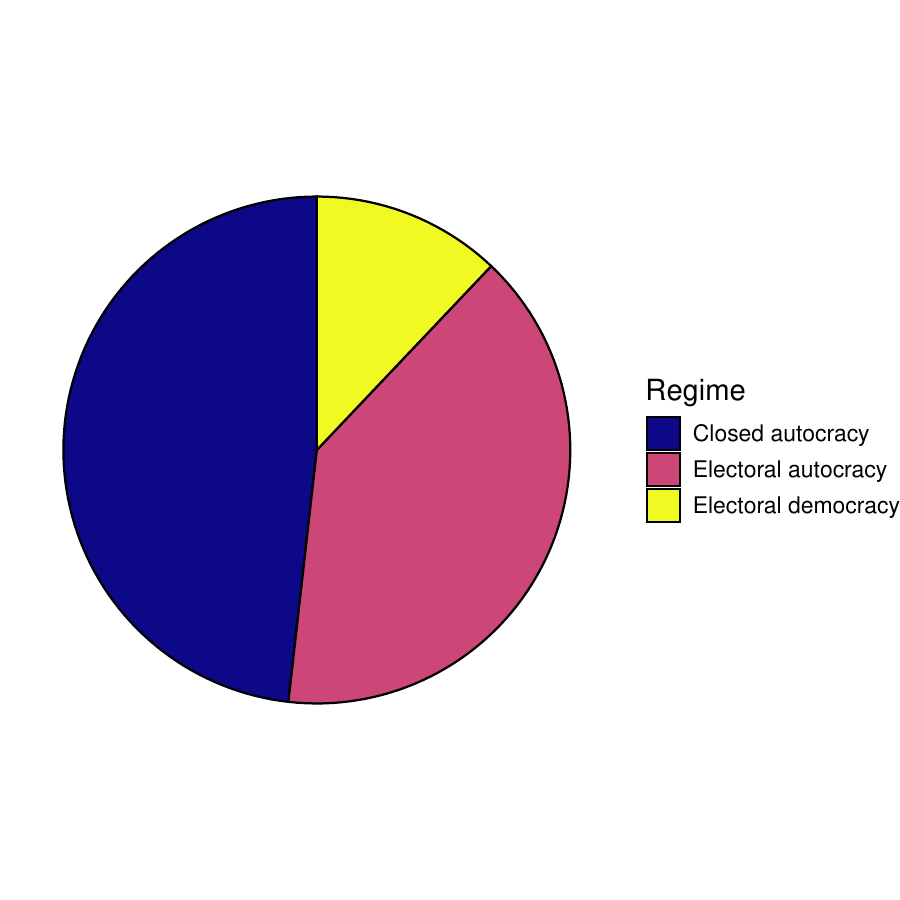}
        \caption{Regime}
        \label{fig:reg}
    \end{subfigure}
    \hfill
    \begin{subfigure}[b]{0.55\textwidth}
        \centering
        \includegraphics[width=\textwidth]{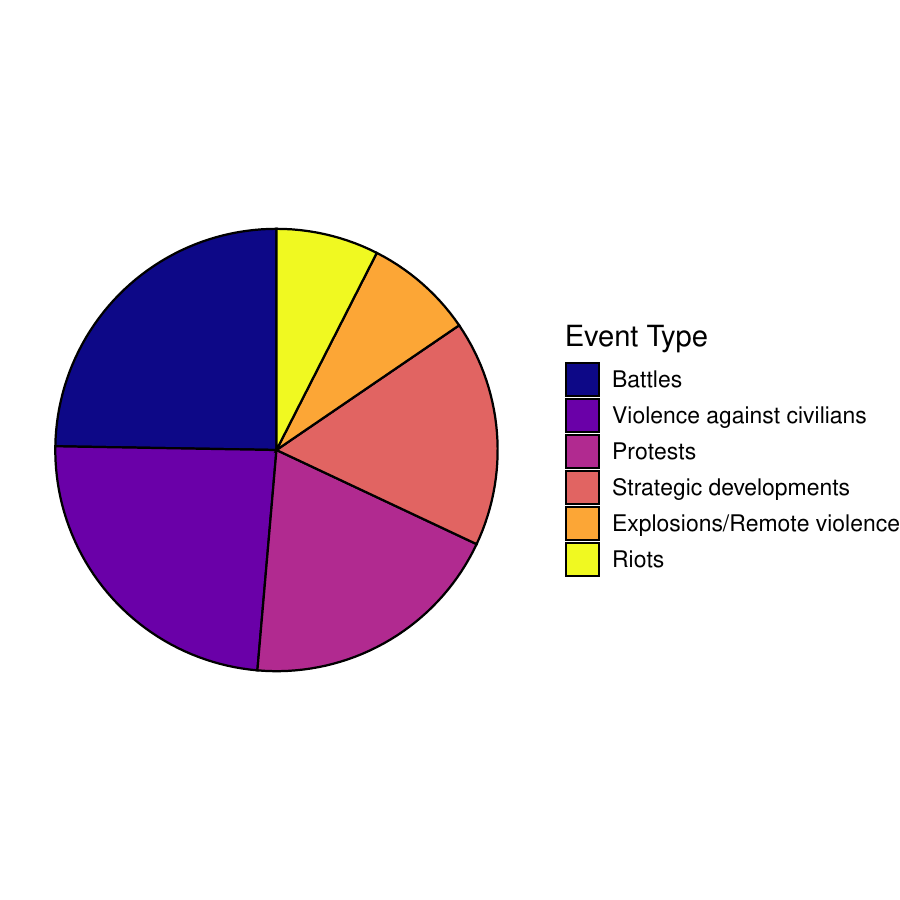}
        \caption{Event type}
        \label{fig:event}
    \end{subfigure}
    \caption{Data distribution by fatality, regime type, and event type}
    \label{fig:smooth_plots}
\end{figure}

\end{document}